\documentclass[pdflatex,sn-mathphys]{sn-jnl}
\usepackage{amsmath,amssymb}
\usepackage{graphicx}
\usepackage{dcolumn}
\usepackage{bm,color}
\usepackage{hyperref}
\usepackage{accents}
\usepackage{amssymb,float}
\usepackage{amsmath}
\usepackage{multirow}
\usepackage{siunitx}
\usepackage{tabularx}
\usepackage{booktabs}
\usepackage{url}

\DeclareSIUnit\parsec{pc}

\newcommand{\udt}[3]{#1^{#2}_{\phantom{#2}#3}}

\newcommand{\dut}[3]{#1_{#2}^{\phantom{#2}#3}}

\newcommand{\lc}[1]{\accentset{\circ}{#1}}

\jyear{2022}%
\begin{document}

\title[Impact of \texorpdfstring{$H_0$}{} priors on \texorpdfstring{$f(T)$}{} late time cosmology]{Impact of \texorpdfstring{$H_0$}{} priors on \texorpdfstring{$f(T)$}{} late time cosmology}


\author[1,2]{Rebecca Briffa,}\email{rebecca.briffa.16@um.edu.mt}
\author[3]{Celia Escamilla-Rivera,} \email{celia.escamilla@nucleares.unam.mx}
\author[1,2]{Jackson Levi Said,} \email{jackson.said@um.edu.mt}
\author[1,2]{Jurgen Mifsud} \email{jurgen.mifsud@um.edu.mt}
\author[1,2]{and Nathan Lee Pullicino} \email{nathan.pullicino.14@um.edu.mt}

\affil[1]{Institute of Space Sciences and Astronomy, University of Malta, Malta, MSD 2080}
\affil[2]{Department of Physics, University of Malta, Malta}
\affil[3]{Instituto de Ciencias Nucleares, Universidad Nacional Aut\'{o}noma de M\'{e}xico, Circuito Exterior C.U., A.P. 70-543, M\'exico D.F. 04510, M\'{e}xico}


\abstract{We present a detailed analysis of the impact of $H_0$ priors from recent surveys in the literature on the late time cosmology of five $f(T)$ cosmological models using cosmic chronometers, the Pantheon data set, and baryonic acoustic oscillation data. In this work, we use three recently reported values of $H_0$ that have contributed to the recent $H_0$ tension problem. We find that these priors have a strong response in these analyses in terms of all the cosmological parameters. In general, our analyses gives much higher values of $H_0$ when considered against equivalent analyses without priors while, by and large, giving lower values of the matter density parameter. We close with a cross-analysis of each of our model, data set and prior combination choices.
}


\keywords{Cosmology, Dark Energy}

\maketitle


\section{\label{sec:intro}Introduction}

The growing pressure from the so-called $H_0$ tension \cite{Bernal:2016gxb,DiValentino:2020zio,DiValentino:2021izs} has prompted a reconsideration of novel approaches to formulating a consistent cosmological model. On the other hand, the standard model of cosmology, the $\Lambda$CDM cosmological model, rests on overwhelming consistency with observational cosmology \cite{Planck:2018vyg}. This is only possible with the inclusion of matter beyond the standard model of particle physics in terms of cold dark matter (CDM) which stabilizes galactic structures \cite{Baudis:2016qwx,Bertone:2004pz} while on larger scales dark energy, through the cosmological constant ($\Lambda$) \cite{Peebles:2002gy,Copeland:2006wr}, produces the acceleration observed in this regime \cite{SupernovaSearchTeam:1998fmf,SupernovaCosmologyProject:1998vns}. While great efforts have been put into the detection of these exotic forms of matter, they remain elusive to direct observations \cite{Gaitskell:2004gd}, and continue to be plagued by foundational issues \cite{Weinberg:1988cp}.

The prospect of a possible observational disparity in $\Lambda$CDM has prompted renewed efforts to determine the current value of the cosmological expansion in order to better assess the degree of inconsistency that the concordance model may be expressing. This parameter characterizes the discrepancy between $\Lambda$CDM independent measurements of the Hubble parameter at current times \cite{Riess:2019cxk,Wong:2019kwg} and its $\Lambda$CDM-based predicted value from early time observations \cite{Planck:2018vyg,DES:2021wwk}. While tip of the red giant branch (TRGB, Carnegie-Chicago Hubble Program) point to a lower $H_0$ tension, the $H_0$ tension has been growing for some time and may only be resolved either by novel observations such as using using gravitational astronomy \cite{Graef:2018fzu,Abbott:2017xzu,Baker:2019nia,2017arXiv170200786A}, or possibly by considering other gravitational contributions to our cosmological model.

There exist many possible modifications of general relativity (GR) that may be further developed into an observationally viable gravitational base for a concordance model for cosmology \cite{Sotiriou:2008rp,Clifton:2011jh,CANTATA:2021ktz}. These are largely built on correction terms to the Einstein-Hilbert action \cite{Faraoni:2008mf,Capozziello:2011et}. Here, gravitational interactions continue to be based on the curvature associated with the Levi-Civita connection, which is the source of curvature in GR \cite{misner1973gravitation,nakahara2003geometry}. However, there is a growing body of work in which torsion rather than curvature is considered as the form in which gravitation is expressed \cite{Bahamonde:2021gfp,Aldrovandi:2013wha,Cai:2015emx,Krssak:2018ywd}, which has produced interesting cosmological models. Teleparallel gravity (TG) incorporates the theories in which the teleparallel connection \cite{Weitzenbock1923,Bahamonde:2021gfp} is used, which expresses this torsion in geometry. The teleparallel connection is curvature-less and satisfies metricity, and so all measures of curvature identically vanish irrespective of the components of the metric. A consequence of this is that the regular Ricci scalar $\lc{R}$ (over-circles represent quantities calculated with the Levi-Civita connection) will be zero when calculated with the teleparallel connection, i.e. $R=0$. By relating both forms of the Ricci scalar, TG produces a torsion scalar $T$ which is dynamically equivalent to GR, called the \textit{Teleparallel equivalent of General Relativity} (TEGR), which differs from the Einstein-Hilbert action by a boundary term $B$. The division between the torsion scalar and boundary terms means that a much larger range of theories that are second-order in derivatives can be formed giving a weaker form of the generalized Lovelock theory \cite{Lovelock:1971yv,Gonzalez:2015sha,Bahamonde:2019shr}.

Using the same rationale as in other modified theories of gravity, such as $f(\lc{R})$ gravity \cite{Sotiriou:2008rp,Faraoni:2008mf,Capozziello:2011et}, TEGR can be directly generalized to $f(T)$ gravity \cite{Ferraro:2006jd,Ferraro:2008ey,Bengochea:2008gz,Linder:2010py,Chen:2010va,Bahamonde:2019zea}. $f(T)$ gravity is a second-order gravitational theory that has shown promise at meeting the observational challenges that are becoming all the more pressing \cite{Cai:2015emx,Farrugia:2016qqe,Finch:2018gkh,Farrugia:2016xcw,Iorio:2012cm,Deng:2018ncg}. In Ref.~\cite{Nesseris:2013jea} both expansion and growth data are used to constrain several prominent models of $f(T)$ gravity, resulting in model parameters that are within 1$\sigma$ of their corresponding $\Lambda$CDM values where specific extended models are considered. The three most promising extended models are then reconsidered in Ref.~\cite{Anagnostopoulos:2019miu} where the most recent observations are considered, and which shows consistency with $\Lambda$CDM. $f(T)$ gravity has also been analyzed in terms of its impact on the CMB power spectrum from gravitational waves where in Ref.~\cite{Nunes:2018evm} the power-law model was explored in this regime. More recently, Ref.~\cite{Benetti:2020hxp} include data from big bang nucleosynthesis to constrain further these three models. 

In recent years there have been a number of cosmology independent measurements of the current value of Hubble parameter, which is the main driver of the growing tension between local, and early-Universe predicted values of $H_0$. In this work, we explore the impact of these priors on five core models in $f(T)$ cosmology. The literature contains several additional works where observational data is used to constrain TG models with various uses of priors in these models. To explore the impact of these values of measurements of $H_0$ we perform several background studies on these models using these various settings in order to better discern the impacts of priors and $f(T)$ models.

TG also produces other interesting theories such as $f(T,B)$ gravity \cite{Bahamonde:2015zma,Bahamonde:2016grb,Paliathanasis:2017flf,Farrugia:2018gyz,Bahamonde:2016cul,Wright:2016ayu,Farrugia:2020fcu,Capozziello:2019msc,Farrugia:2018gyz,Escamilla-Rivera:2019ulu} where the boundary term plays a more active role, as well as $f(T,T_G)$ where $T_G$ is the Gauss-Bonnet invariant \cite{Kofinas:2014owa,Kofinas:2014daa,delaCruz-Dombriz:2017lvj,delaCruz-Dombriz:2018nvt}, and numerous scalar-tensor theories such as \cite{Bahamonde:2019shr,Bahamonde:2019ipm,Bahamonde:2020cfv}. We first describe the technical details of TG and its modification to $f(T)$ gravity in Sec.~\ref{sec:intro_f_T}, while in Sec.~\ref{sec:obs_data} we describe the data sets we use in our Markov chain Monte Carlo (MCMC) implementation. Our core results are presented in Sec.~\ref{sec:models} where we also introduce the five models under consideration. We close with a summary of our results and their place against similar studies in the literature in Sec.~\ref{sec:conc}.

\section{\label{sec:intro_f_T}\texorpdfstring{$f(T)$}{ft} FLRW Cosmology}

TG recasts the curvature of GR and its modifications \cite{Clifton:2011jh} with a torsional geometric framework \cite{Hohmann:2019nat} which is built on the exchange of the Levi-Civita connection with the teleparallel connection. This then envelopes into a gravitational theory that embodies torsion from its foundations \cite{Bahamonde:2021gfp}.

The source of curvature in GR is the Levi-Civita connection $\udt{\lc{\Gamma}}{\sigma}{\mu\nu}$ (over-circles are used throughout to denote quantities determined using the Levi-Civita connection) rather than the metric tensor, which acts as the dynamical variable of the theory but actually only quantifies the amount of geometric deformation. TG characterizes gravitation as torsion through its associated teleparallel connection $\udt{\Gamma}{\sigma}{\mu\nu}$ which satisfies metricity but which is curvature-less \cite{Hayashi:1979qx,Aldrovandi:2013wha}. This realization means that all the curvature-based quantities will identically vanish for the teleparallel connection such as the teleparallel Riemann tensor (the regular Levi-Civita connection Riemann tensor naturally does not vanish). In this background, TG requires an entirely new formulation of gravitational tensors on which to build theories (see reviews in Refs. \cite{Krssak:2018ywd,Cai:2015emx,Aldrovandi:2013wha}).

The most direct way of formulating teleparallel theories of gravity is through the tetrad $\udt{e}{A}{\mu}$ (and its inverses $\dut{E}{A}{\mu}$) which replaces the metric tensor as the fundamental dynamical variable in TG theories through
\begin{align}\label{metric_tetrad_rel}
    g_{\mu\nu}=\udt{e}{A}{\mu}\udt{e}{B}{\nu}\eta_{AB}\,,& &\eta_{AB} = \dut{E}{A}{\mu}\dut{E}{B}{\nu}g_{\mu\nu}\,,
\end{align}
where Latin indices represent coordinates on the tangent space while Greek indices represent coordinates on the general manifold \cite{Cai:2015emx}. In GR, tetrads are largely suppressed in their usage, as an example, they are used to describe spinors \cite{Chandrasekhar:1984siy}. Naturally, tetrads have to satisfy orthogonality conditions
\begin{align}
    \udt{e}{A}{\mu}\dut{E}{B}{\mu}=\delta^A_B\,,&  &\udt{e}{A}{\mu}\dut{E}{A}{\nu}=\delta^{\nu}_{\mu}\,,
\end{align}
for internal consistency.

The teleparallel connection can be defined as \cite{Weitzenbock1923,Krssak:2018ywd}
\begin{equation}
    \udt{\Gamma}{\sigma}{\nu\mu} := \dut{E}{A}{\sigma}\left(\partial_{\mu}\udt{e}{A}{\nu} + \udt{\omega}{A}{B\mu}\udt{e}{B}{\nu}\right)\,,
\end{equation}
where $\udt{\omega}{A}{B\mu}$ is a flat spin connection which is responsible for incorporating the local Lorentz transformation invariance of the theory (which arises explicitly due to the appearance of the tangent space indices). This can be contrasted with GR where the spin connections (associated with their tetrads) are not flat \cite{misner1973gravitation}. In TG, tetrad-spin connection pairs represent gravitational and local degrees of freedom respectively and both contribute to a system's equations of motion. Similar to the way that the Levi-Civita connection builds up to the Riemann tensor, the teleparallel connection can be used to describe a torsion tensor \cite{Hayashi:1979qx}
\begin{equation}
    \udt{T}{\sigma}{\mu\nu}  :=2\udt{\Gamma}{\sigma}{[\nu\mu]}\,,
\end{equation}
where square brackets denote an antisymmetric operator, and where $\udt{T}{\sigma}{\mu\nu}$ represents the gauge field strength of gravity in TG \cite{Aldrovandi:2013wha}. The torsion tensor is covariant under both diffeomorphisms and local Lorentz transformations. By taking suitable contractions of the torsion tensor, a torsion scalar can be written down as \cite{Krssak:2018ywd,Cai:2015emx,Aldrovandi:2013wha,Bahamonde:2021gfp}
\begin{equation}
    T:=\frac{1}{4}\udt{T}{\alpha}{\mu\nu}\dut{T}{\alpha}{\mu\nu} + \frac{1}{2}\udt{T}{\alpha}{\mu\nu}\udt{T}{\nu\mu}{\alpha} - \udt{T}{\alpha}{\mu\alpha}\udt{T}{\beta\mu}{\beta}\,,
\end{equation}
which can be arrived at either by demanding that $T$ be equivalent to the Ricci scalar (up to a boundary term), or by interpreting TG as a gauge theory of translations, which then naturally leads to this form of the torsion scalar. The Ricci scalar is dependent only on the Levi-Civita connection, and similarly, the torsion scalar is entirely dependent on the teleparallel connection.

Theories based on the Levi-Civita connection produce the Ricci scalar $\lc{R}$ among other measures of curvature. By exchanging this with the teleparallel connection, this form of the Ricci scalar will identically vanish, meaning $R\equiv0$ (where we emphasize that $R = R(\udt{\Gamma}{\sigma}{\mu\nu})$ and $\lc{R}=\lc{R}(\udt{\lc{\Gamma}}{\sigma}{\mu\nu})$). In terms of gravitational scalars, the torsion and Ricci scalar are equivalent up to a boundary term $B$, which can be represented through the relation \cite{Bahamonde:2015zma,Farrugia:2016qqe}
\begin{equation}\label{LC_TG_conn}
    R=\lc{R} + T - B = 0\,.
\end{equation}
This is the base of the TEGR action, which is thus guaranteed to feature identical equations of motion as the Einstein-Hilbert action.

Following the same reasoning as the many extensions to GR, such as $f(\lc{R})$ gravity \cite{DeFelice:2010aj,Capozziello:2011et}, TEGR can be arbitrarily extended to $f(T) = -T + \mathcal{F}(T)$ gravity by raising the TEGR action to \cite{Ferraro:2006jd,Ferraro:2008ey,Bengochea:2008gz,Linder:2010py,Chen:2010va,RezaeiAkbarieh:2018ijw}
\begin{equation}\label{f_T_ext_Lagran}
    \mathcal{S}_{\mathcal{F}(T)}^{} =  \frac{1}{2\kappa^2}\int \mathrm{d}^4 x\; e\left(-T + \mathcal{F}(T)\right) + \int \mathrm{d}^4 x\; e\mathcal{L}_{\text{m}}\,,
\end{equation}
where $\kappa^2=8\pi G$, $\mathcal{L}_{\text{m}}$ is the matter Lagrangian, and $e=\det\left(\udt{e}{a}{\mu}\right)=\sqrt{-g}$ is the tetrad determinant. The TEGR limit will then be described by $\mathcal{F}(T) \rightarrow 0$, while $\Lambda$CDM is recovered when the arbitrary Lagrangian tends to a constant $\Lambda$ value. The most important difference, in terms of the dynamical equations, between $\mathcal{F}(\lc{R})$ and $\mathcal{F}(T)$ is that the total divergence term $B$ \eqref{LC_TG_conn} is no longer a boundary term in $\mathcal{F}(\lc{R})=\mathcal{F}(-T+B)$ gravity. This is the reason why $\mathcal{F}(\lc{R})$ equations of motion become fourth-order while $f(T)$ remains second-order, as in TEGR. This is advantageous for several reasons such as being naturally Gauss-Ostrogadsky ghost free \cite{Aldrovandi:2013wha} and being more amenable to numerical approaches.

The field equations for $f(T)$ gravity can then be written as
\begin{align}\label{ft_FEs}
    \dut{W}{a}{\mu} := e^{-1} &\partial_{\nu}\left(e\dut{E}{a}{\rho}\dut{S}{\rho}{\mu\nu}\right)\left(-1 + \mathcal{F}_T\right) - \dut{E}{a}{\lambda} \udt{T}{\rho}{\nu\lambda}\dut{S}{\rho}{\nu\mu} \left(-1 + \mathcal{F}_T\right) + \frac{1}{4}\dut{E}{a}{\mu}\left(-T + \mathcal{F}(T)\right) \nonumber\\
    & + \dut{E}{a}{\rho}\dut{S}{\rho}{\mu\nu}\partial_{\nu}\left(T\right)\mathcal{F}_{TT}  + \dut{E}{b}{\lambda}\udt{\omega}{b}{a\nu}\dut{S}{\lambda}{\nu\mu}\left(-1 + \mathcal{F}_T\right) = \kappa^2 \dut{E}{a}{\rho} \dut{\Theta}{\rho}{\mu}\,,
\end{align}
where subscripts denote derivatives ($\mathcal{F}_T=\partial \mathcal{F}/\partial T$ and  $\mathcal{F}_{TT}=\partial^2 \mathcal{F}/\partial T^2$), and $\dut{\Theta}{\rho}{\nu}$ is the regular energy-momentum tensor. The separate tetrad and spin connection variations produce the field equations \cite{Bahamonde:2021gfp}
\begin{equation}
    W_{(\mu\nu)} = \kappa^2 \Theta_{\mu\nu}\,, \quad \text{and} \quad W_{[\mu\nu]} = 0\,,
\end{equation}
which represent the degrees of freedom associated with the tetrad and spin connection, respectively. For any metric ansatz a unique tetrad-spin connection pair exists such that the local frame is compatible with all the spin connection components vanishing, which is called the Weitzenb\"{o}ck gauge \cite{Krssak:2018ywd,Bahamonde:2021gfp}. In this setting, $W_{[\mu\nu]}$ vanishes taking a select choice of tetrad components, while still satisfying the metric equations in Eq.~\eqref{metric_tetrad_rel}.

In this work we explore the cosmology of a flat homogeneous and isotropic Universe which can be represented by the tetrad  \cite{Krssak:2015oua,Tamanini:2012hg}
\begin{equation}
    \udt{e}{A}{\mu} = \text{diag}\left(1,\,a(t),\,a(t),\,a(t)\right)\,,
\end{equation}
where $a(t)$ is the scale factor in cosmic time $t$, and which was shown to universally satisfy the Weitzenb\"{o}ck gauge conditions in Ref.~\cite{Hohmann:2019nat}. This reproduces, through Eq.~\eqref{metric_tetrad_rel}, the regular flat Friedmann--Lema\^{i}tre--Robertson--Walker (FLRW) metric \cite{misner1973gravitation}
\begin{equation}\label{FLRW_metric}
     \mathrm{d}s^2 = \mathrm{d}t^2 - a^2(t) \left(\mathrm{d}x^2+\mathrm{d}y^2+\mathrm{d}z^2\right)\,.
\end{equation} 

Taking the standard definition of Hubble parameter as $H=\dot{a}/a$ where over-dots refer to derivatives with respect to cosmic time, the equivalence in Eq.~\eqref{LC_TG_conn}. In this setting, we find $T = -6 H^2$ and $B = -6\left(3H^2 + \dot{H}\right)$, which produces the regular Ricci scalar through
\begin{equation}
    \lc{R} = -T + B = -6\left(2H^2 + \dot{H}\right)\,.
\end{equation}
The Friedmann equations can then be written down as \cite{Bahamonde:2021gfp}
\begin{align}
    H^2 + \frac{T}{3}\mathcal{F}_T - \frac{\mathcal{F}}{6} &= \frac{\kappa^2}{3}\rho\,,\label{eq:Friedmann_1}\\
    \dot{H}\left(1 - \mathcal{F}_T - 2T\mathcal{F}_{TT}\right) &= -\frac{\kappa^2}{2} \left(\rho + p \right)\label{eq:Friedmann_2}\,,
\end{align}
where we denote the energy density and pressure of the total matter sector by $\rho$ and $p$, respectively.

\section{\label{sec:obs_data}Observational Data}

We here present the observational data sets which will be considered in the below analyses. For our baseline data set, we consider $H(z)$ data along with a Supernovae type Ia (SNIa) compilation data set.

\begin{itemize}
    \item For the Hubble parameter data, we adopt thirty--one data points which were inferred via the cosmic chronometers (CC) technique. Such a technique enables us to directly derive information about the Hubble function at several redshifts, up to $z\lesssim2$. Since the adopted CC data is primarily based on measurements of the age difference between two passively--evolving galaxies that formed at the same time but are separated by a small redshift interval (from which one can compute $\Delta z/\Delta t$), CC were found to be more reliable than any other method based on an absolute age determination for galaxies \cite{Jimenez:2001gg}. Our adopted CC data points were compiled from Refs. \cite{2014RAA....14.1221Z,Jimenez:2003iv,Moresco:2016mzx,Simon:2004tf,2012JCAP...08..006M,2010JCAP...02..008S,Moresco:2015cya}, which are independent of the Cepheid distance scale and from any cosmological model, although they are dependent on the modelling of stellar ages, which is based on robust stellar population synthesis techniques (see, for instance, Refs. \cite{Gomez-Valent:2018hwc,Lopez-Corredoira:2017zfl,Lopez-Corredoira:2018tmn,Verde:2014qea,2012JCAP...08..006M,Moresco:2016mzx} for analyses related to CC systematics). The corresponding $\chi^2_H$ estimator is given by 
\begin{equation}
    \chi^2_H(\Theta)=\sum_{i=1}^{31}\frac{\left(H(z_i,\,\Theta)-H_\mathrm{obs}(z_i)\right)^2}{\sigma_H^2(z_i)}\,,
\end{equation}
where $H(z_i,\Theta)$ are the theoretical Hubble parameter values at redshift $z_i$ with model parameters $\Theta$, $H_\mathrm{obs}(z_i)$ are the corresponding measured values of the Hubble parameter at $z_i$ with observational error of $\sigma_H(z_i)$.

\item The other component of our baseline data set consists of the Pantheon compilation of 1048 SNIa relative luminosity distance measurements spanning the redshift range of $0.01<z<2.3$ \cite{Scolnic:2017caz}. Henceforth, we will be denoting the Pantheon SNIa compilation by SN. The publicly available release of the SN catalog provides SNIa magnitudes corrected for systematic effects, including  the stretch of the light--curve, the color at maximum brightness and the stellar mass of the host galaxy. Since the apparent magnitude of each SNIa needs to be calibrated via an arbitrary fiducial absolute magnitude $M$, we will be considering $M$ as a nuisance parameter in our MCMC analyses. This can be implemented through the use of the theoretical values of the distance moduli
\begin{equation}
    \mu(z_i,\,\Theta)=5\log_{10}\left[D_L(z_i,\,\Theta)\right]+M\,,
\end{equation}
at redshift $z_i$ via the corresponding computation of the luminosity distance
\begin{equation}
    D_L(z_i,\Theta)=c\,(1+z_i)\int_0^{z_i}{\frac{\mathrm{d}z'}{H(z',\,\Theta)}}\,,
\end{equation}
where $c$ is the speed of light, and the nuisance parameter $M$ encodes the Hubble constant which has to be marginalized over in the MCMC analyses. The associated $\chi^2_\mathrm{SN}$ is specified by \cite{2011ApJS..192....1C}
\begin{equation}
    \chi_{\text{SN}}^2(\Theta) = \left(\Delta\mu(z_i,\,\Theta)\right)^{T} C_{\text{SN}}^{-1}\, \Delta\mu(z_i,\,\Theta)+\ln\left({\frac{S}{2\pi}}\right)-\frac{k^2(\Theta)}{S}\,,
\end{equation}
where $C_{\text{SN}}^{}$ is the total covariance matrix, $S$ is the sum of all the components of $C_{SN}^{-1}$, while $k$ is given by
\begin{equation}
    k(\Theta)={\left(\Delta\mu(z_i,\,\Theta)\right)^{T}\cdotp C_{\text{SN}}^{-1}}\,,
\end{equation}
with $\Delta\mu(z_i,\,\Theta)=\mu(z_i,\,\Theta)-\mu_{\text{obs}}(z_i)$.

\item We further consider a joint baryon acoustic oscillation (BAO) data set composed of independent data points. The BAO data set incorporates the SDSS Main Galaxy Sample measurement at $z_{\mathrm{eff}}=0.15$ \cite{Ross:2014qpa}, the six--degree Field Galaxy Survey measurement at $z_{\mathrm{eff}}=0.106$ \cite{2011MNRAS.416.3017B}, and the BOSS DR11 quasar Lyman--$\alpha$ measurement at $z_{\mathrm{eff}}=2.4$ \cite{Bourboux:2017cbm}. We further consider the angular diameter distances and $H(z)$ measurements of SDSS--IV eBOSS DR14 quasar survey at $z_{\mathrm{eff}}=\{0.98,\,1.23,\,1.52,\,1.94\}$ \cite{Zhao:2018gvb}, along with the SDSS--III BOSS DR12 consensus BAO measurements of the Hubble parameter and the corresponding comoving angular diameter distances at $z_{\mathrm{eff}}=\{0.38,\,0.51,\,0.61\}$ \cite{Alam:2016hwk}, where in these two BAO data sets we consider the full covariance matrix in our MCMC analyses. 
For the considered BAO data sets we computed the Hubble distance $D_H(z)$, comoving angular diameter distance $D_M(z)$, and volume--average distance $D_V(z)$, which are respectively specified by
\begin{equation}
    D_H(z)=\frac{c}{H(z)}\,,\;\;D_M(z)=(1+z)D_A(z)\,,\;\;D_V(z)=\left[(1+z)^2D_A^2(z)\frac{z}{H(z)}\right]^{1/3}\,,
\end{equation}
with $D_A(z)=(1+z)^{-2}D_L(z)$ being the angular diameter distance. In order to use the reported BAO results in our MCMC analyses, we had to consider the corresponding combination of parameters $\mathcal{G}(z_i)=\{D_V(z_i)/r_s(z_d),\allowbreak\,r_s(z_d)/D_V(z_i),\allowbreak\,D_H(z_i),\allowbreak\,D_M(z_i)(r_{s,\mathrm{fid}}(z_d)/r_s(z_d)),\allowbreak\,H(z_i)(r_s(z_d)/r_{s,\mathrm{fid}}(z_d)),\allowbreak\,D_A(z_i)(r_{s,\mathrm{fid}}(z_d)/r_s(z_d))\}$, for which we had to compute the comoving sound horizon $r_s(z)$ at the end of the baryon drag epoch at redshift $z_d\approx1059.94$ \cite{Planck:2018vyg}, such that
\begin{equation}
    r_s(z)=\int_z^\infty\frac{c_s(\tilde{z})}{H(\tilde{z})}\,\mathrm{d}z=\frac{1}{\sqrt{3}}\int_0^{1/(1+z)}\frac{\mathrm{d}a}{a^2H(a)\sqrt{1+\left[3\Omega_{b,0}/(4\Omega_{\gamma,0})\right]a}}\,,
\end{equation}
where we have adopted $\Omega_{b,0}=0.02242$ \cite{Planck:2018vyg}, $T_{0}=2.7255\,\mathrm{K}$ \cite{2009ApJ...707..916F}, and a fiducial value of $r_{s,\mathrm{fid}}(z_d)=147.78\,\mathrm{Mpc}$. The corresponding $\chi^2_\mathrm{BAO}(\Theta)$ is specified by
\begin{equation}
    \chi^2_\mathrm{BAO}(\Theta)=(\Delta\mathcal{G}(z_i,\Theta))^T\,C^{-1}_\mathrm{BAO}\,\Delta\mathcal{G}(z_i,\Theta)\,,
\end{equation}
where $\Delta\mathcal{G}(z_i,\Theta)=\mathcal{G}(z_i,\Theta)-\mathcal{G}_\mathrm{obs}(z_i)$, and $C_\mathrm{BAO}$ is the covariance matrix of all the considered BAO observations.
\end{itemize}

Further to the above data sets, we will also be analysing the impact of an $H_0$ prior value on our $f(T)$ model parameter constraints. We will be considering the latest SH0ES local estimate \cite{Riess:2019cxk} of $H_0= 74.22 \pm 1.82 \,{\rm km\, s}^{-1} {\rm Mpc}^{-1}$ (R19) based on SN in the Hubble flow, the H0LiCOW Collaboration's \cite{Wong:2019kwg} measurement which relies on strong lensing from quasars and has a value of $73.3^{+1.7}_{-1.8} \,{\rm km\, s}^{-1} {\rm Mpc}^{-1}$ (HW), and finally the measurement using the tip of the red giant branch (TRGB) as a standard candle with $H_0=69.8 \pm 1.9 \,{\rm km\, s}^{-1} {\rm Mpc}^{-1}$ \cite{Freedman:2019jwv}. While other measurements exist (see, for instance, Ref.~\cite{Abbott:2017xzu}), the aforementioned measurements are the most representative model--independent local values which have been exhaustively studied in the literature.

\section{\label{sec:models}Constraints of \texorpdfstring{$f(T)$}{fT} Cosmological Models}

We take five models within the $f(T)$ gravity framework in which to probe the impact of recent cosmology model independent measurements of $H_0$. These have gained prominence in the literature and are frequently studied since they are reported to mirror very well our cosmological history. We analyze the impact that these priors have on the resulting parameter values for the data sets described in Sec.~\ref{sec:obs_data}. In this context, we try to separate the impacts between the choice of prior on $H_0$ and the $f(T)$ model under consideration.

In order to better evaluate the performance of each $f_i$CDM model against $\Lambda$CDM, we compute the Akaike information criterion (AIC) \cite{Akaike:1974} to assess which model ultimately best supports the data. The AIC is defined as
\begin{equation}
    \text{AIC} = -2\ln L_{\rm max} + 2n\,,
\end{equation}
where $L_{\rm max}$ is the maximum value of the likelihood function for each model and data set together with prior combination, and $n$ is the number of parameters involved in the estimation routine. In practice, lower values of AIC indicate better performance against observational data set combinations. On the other hand, we also consider the Bayesian information criterion (BIC) \cite{10.1214/aos/1176344136} which is closely related to AIC and defined as
\begin{equation}
    \text{BIC} = -2\ln L_{\rm max} + n\ln m\,,
\end{equation}
where $m$ is the sample size of the observational data combination. This variant of the AIC also imposes a penalty for models having a large number of parameters to estimate, which is more impactful in BIC. The explicit appearance of the number of points in a data set also makes the BIC an arguably better measure of the performance of AIC against observational data.

\subsection{\texorpdfstring{$f_1(T)$}{}CDM Model}
The power-law model \cite{Bengochea:2008gz} was first explored due to its ability to produce an accelerated late-time Universe. This model is expressed as
\begin{equation}\label{eq:f1}
    \mathcal{F}_1 (T) = \alpha_1 \left(-T\right)^{b_1}\,,
\end{equation}
where $\alpha_1$ and $b_1$ are constants. By evaluating the Friedmann equation \eqref{eq:Friedmann_1} at current times, we can obtain
\begin{equation}
    \alpha_1 = \left(6H_0^2\right)^{1-b_1} \frac{1-\Omega_{M_0}}{1 - 2b_1}\,,
\end{equation}
where $\Omega_{M_0} = \Omega_{m_0} + \Omega_{r_0}$, and where $\Omega_{m_0}$ and $\Omega_{r_0}$ are the density parameters for matter and radiation at current times, respectively. This makes $b_1$ the only new model parameter for $f_1$CDM rather than the superficial addition of two parameters shown in Eq.~(\ref{eq:f1}).

By defining the normalised Hubble parameter $E(z) := H(z)/H_0$, we can write the Friedmann equation \ref{eq:Friedmann_1} for this model as
\begin{equation}
    E^2(z) = \Omega_{m_0} \left(1+z\right)^3 + \Omega_{r_0}\left(1+z\right)^4 + \left(1 - \Omega_{m_0} - \Omega_{r_0}\right) E^{2b_1}(z)\,,
\end{equation}
which reduces to the $\Lambda$CDM model for $b_1 = 0$, and to the Dvali, Gabadadze and Porrati (DGP) \cite{Dvali:2000hr,Barcenas-Enriquez:2018ili} model for $b_1 = 1/2$. For $b_1 = 1$, the additional component in the Friedmann equation produces a rescaled gravitational constant term in the density parameters, i.e. this is the pure GR limit. This gives an upper bound such that $b_1 < 1$ for an accelerating Universe.

\begin{figure}[h]
\begin{minipage}{0.5\textwidth}
    \includegraphics[width = 1\textwidth]{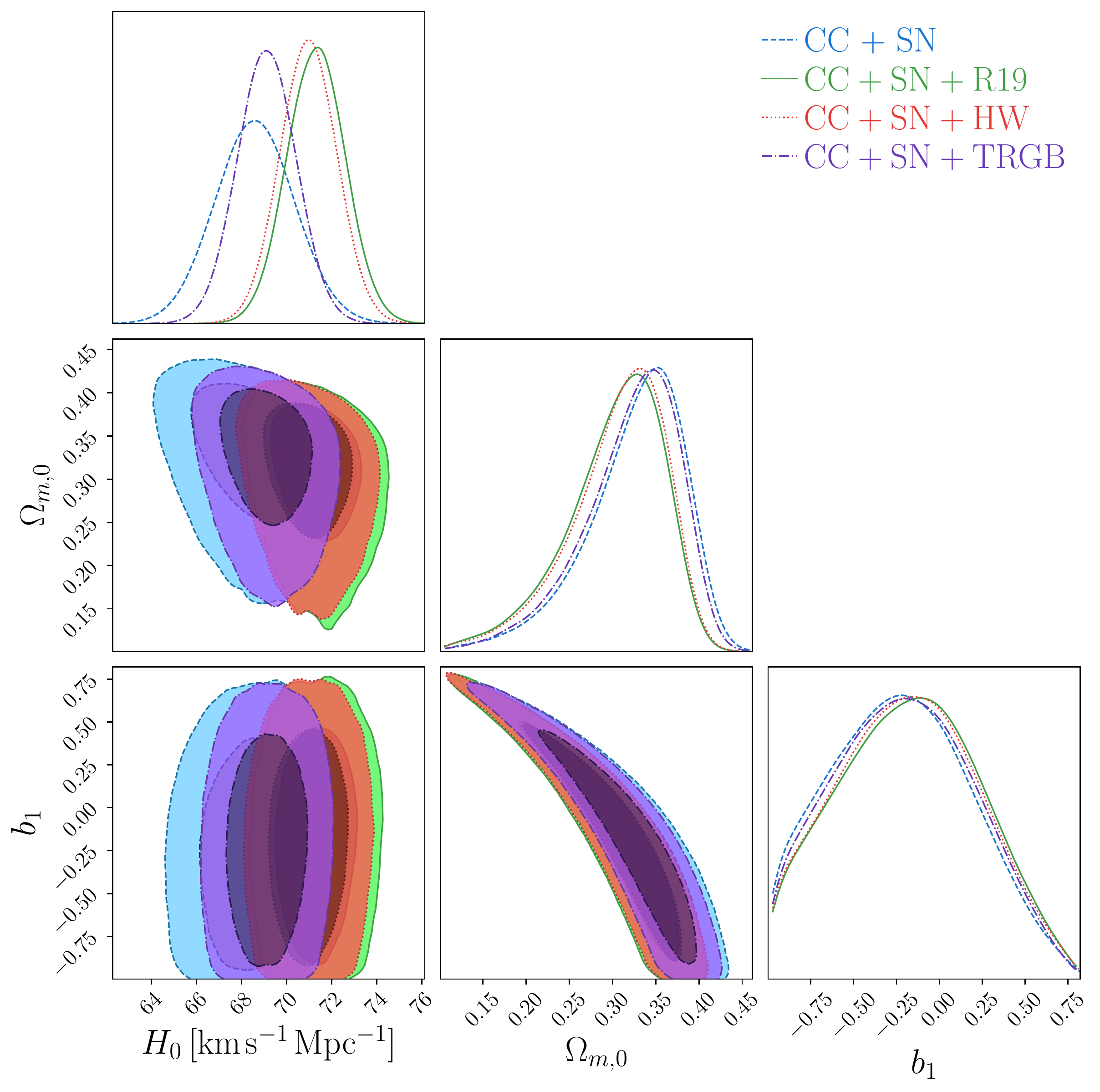}
\end{minipage}
\begin{minipage}{0.5\textwidth}
    \includegraphics[width = 1\textwidth]{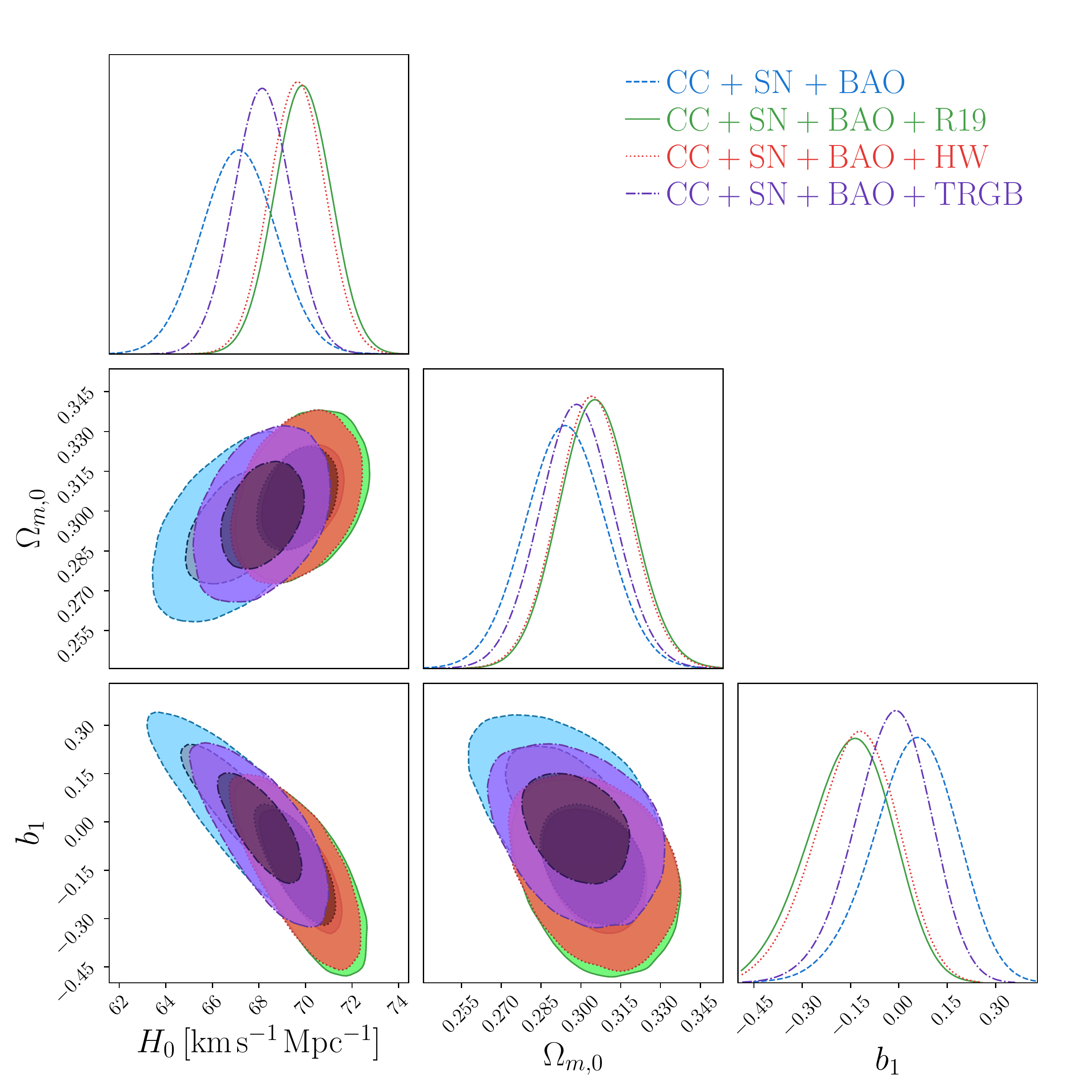}
\end{minipage}
    
    \caption{\textit{Left:} Confidence levels (C.L) and posteriors for the $f_1$CDM (Power Law) model (\ref{eq:f1}) using CC+SN data along the $H_0$ priors: R19 (green color), HW (red color) and TRGB (purple color), are respectively shown. \textit{Right:} C.L and posteriors for the Power Law model (\ref{eq:f1}) using CC+SN+BAO data are shown using the same prior colors.}
    \label{fig:f1CCSN}
\end{figure}

\begin{table}
\tiny
    \centering
    \caption{Results for the $f_1$CDM (Power Law) model (\ref{eq:f1}). First column: data sets used to constrain the model, including the $H_0$ priors. Second column: $H_0$ values derived from the analysis. Third column: Constrained $\Omega_{m,0}$. Fourth column: Best fit $b_1$ values. Fifth column: Nuisance parameter, $M$. Sixth column: $\chi^{2}_{\text{min}}$. From seventh up to tenth column: AIC and BIC with their respective differences with $\Lambda$CDM within the corresponding data set and prior combination (as shown in Appendix~\ref{sec:app}).}
    \label{tab:PLM}
    \begin{tabular}{>{\centering}m{0.19\textwidth}>{\centering}m{0.1\textwidth}>{\centering}m{0.1\textwidth}>{\centering}m{0.085\textwidth}>{\centering}m{0.115\textwidth}>{\centering}m{0.04\textwidth}>{\centering}m{0.04\textwidth}>{\centering}m{0.04\textwidth}>{\centering}m{0.02\textwidth}>{\centering\arraybackslash}m{0.02\textwidth}}
        \hline
		\rule{0pt}{1.2 \baselineskip}Data Sets &  $H_0$[\si{\km/ \s / \mega \parsec}] & $\Omega_{m,0}$ & $b_1$ & $M$ & $\chi^{2}_\mathrm{min}$ & AIC & BIC & $\Delta$AIC & $\Delta$BIC \\ [4pt]
		\hline
		\rule{0pt}{1.2 \baselineskip}CC + SN & $68.5\pm 1.8$ & $0.350^{+0.045}_{-0.064}$ & $-0.22^{+0.41}_{-0.48}$ & $-19.390^{+0.053}_{-0.055}$& 1040.94 & 1048.94 & 1068.88 & 1.45 & 6.43 \\ [4pt]
		$ \mathrm{CC + SN + R19}$ & $71.3^{+1.3}_{-1.4}$ & $0.326^{+0.045}_{-0.065}$ & $-0.13^{+0.40}_{-0.50}$ & $-19.314^{+0.039}_{-0.038}$& 1045.83 & 1053.83 & 1073.77 & 1.51 & 6.50\\ [4pt]
		$\mathrm{CC+ SN + HW}$ & $71.0\pm 1.3$ & $0.329^{+0.045}_{-0.062}$ & $-0.16^{+0.41}_{-0.48}$ & $-19.324^{+0.038}_{-0.037}$ &  1044.50 & 1052.50 & 1072.44 & 1.51 & 6.50 \\ [4pt]
		$\mathrm{CC + SN + TRGB}$ & $69.1^{+1.4}_{-1.3}$ & $0.344^{+0.045}_{-0.063}$ & $-0.20^{+0.42}_{-0.47}$ & $-19.375\pm 0.040$ & 1041.55 & 1049.55 & 1069.49 & 1.87 & 6.85 \\ [4pt]
		\hline
		\rule{0pt}{1.2 \baselineskip}CC + SN + BAO & $67.1\pm 1.6$ & $0.294\pm 0.015$ & $0.06\pm 0.13$ &  $-19.435\pm 0.047$ & 1057.13 & 1065.13 & 1085.13 & 1.68 & 6.68 \\ [4pt]
		$\mathrm{CC + SN + BAO + R19}$ & $69.9\pm 1.2$ & $0.305^{+0.014}_{-0.013}$ & $-0.14^{+0.12}_{-0.13}$ &  $-19.359^{+0.035}_{-0.034}$ & 1066.87 & 1074.87 & 1094.87 & 0.56 & 5.56\\ [4pt]
		$\mathrm{CC+ SN + BAO + HW}$ & $69.7\pm 1.2$ & $0.304^{+0.014}_{-0.012}$ & $-0.12^{+0.12}_{-0.13}$ & $-19.366^{+0.035}_{-0.033}$ & 1064.92 & 1072.92 & 1086.92 & 0.89 & 5.89\\ [4pt]
		$\mathrm{CC + SN + BAO + TRGB}$ & $68.1\pm 1.2$ & $0.298\pm 0.014$ & $-0.01^{+0.11}_{-0.12}$  &  $-19.407\pm 0.036$ & 1058.56 & 1066.56 & 1086.56 & 2.00 & 7.00\\ [4pt]
		\hline
    \end{tabular}
\end{table}

In Fig.~\ref{fig:f1CCSN}, the posteriors and confidence regions are shown for both CC+SN and CC+SN+BAO data set combinations. In these plots, we also show the results for each of the priors on $H_0$ which are described in Sec.~\ref{sec:obs_data}. Immediately, one notices that the biggest impact of having a prior is to raise the value of $H_0$ in the results from the scenario in which there is no prior. A similar effect occurs for the $\Omega_{\rm m,0}$ parameter but this is less pronounced.

To see the precision values, we give specific results in Table~\ref{tab:PLM} where it becomes clear that the highest value of the Hubble constant is achieved for the CC+SN data set with an R19 prior ($H_0=71.3^{+1.3}_{-1.4} \si{\km \s^{-1} {\mega \parsec}^{-1}}$) which occurs due to early Universe impacts on BAO data having the effect of reducing the value of the $H_0$ parameter while R19 being the highest prior produces a shift to higher values in the MCMC runs. In tandem, the value of $\Omega_{\rm m,0}$ reaches a minimum for the CC+SN data set for the R19 prior, which tallies with having a large value of $H_0$ since most of the energy in this scenario will appear as an effective dark energy. Following a similar reasoning, the lowest value of the Hubble constant appears for the no prior ($H_0=67.1\pm 1.6 \si{\km \s^{-1} {\mega \parsec}^{-1}}$) scenario in which CC+SN+BAO data set is selected. Here, the combination of BAO data and the impact of not assuming a prior on $H_0$ brings the value of the Hubble constant to its lowest point. Conversely, this raises the value of $\Omega_{\rm m,0}$ to its highest value for the CC+SN+BAO data set.

In all cases, the value of the $b_1$ parameter is found to be within 1$\sigma$ of their corresponding $\Lambda$CDM value except for the CC+SN+BAO with an R19 prior run where this moves to being within 2$\sigma$ of the corresponding $\Lambda$CDM value. Coincidentally, from the array of all the MCMC runs, this run also turns out to give the lowest values of both $\Delta$AIC and $\Delta$BIC in comparison to the respective $\Lambda$CDM run which is interesting since this becomes the favoured of model of the group of runs. On the other hand, the $b_4$ parameter may effect cosmic perturbations giving an intriguing division between background and perturbation behaviour in terms of model parameters.

These results are within the confidence regions of previous studies in the literature \cite{Nesseris:2013jea,2018ApJ...855...89X,Benetti:2020hxp,Wang:2020zfv} for the closest prior value. We nuance these results in the literature with more in-depth analysis of the impact of other priors on the eventual cosmological parameter values.

\subsection{\texorpdfstring{$f_2(T)$}{}CDM Model}
The Linder model \cite{Linder:2010py} was designed to produce late-time accelerated expansion and is described by
\begin{equation}\label{eq:f2}
    \mathcal{F}_2 (T) = \alpha_2 T_0 \left(1 - \text{Exp}\left[-b_2\sqrt{T/T_0}\right]\right)\,,
\end{equation}
where $\alpha_2$ and $b_2$ are constants, and where $T_0 = T\vert_{t=t_0} = 6H_0^2$. Evaluating at current times, the Friedmann equation relates these constants through
\begin{equation}
    \alpha_2 = \frac{1-\Omega_{M_0}}{\left(1 + b_2\right)e^{-b_2} - 1}\,,
\end{equation}
which makes $b_2$ the new parameter for the $f_2$CDM model. The corresponding Friedmann equation can then be written as
\begin{equation}
    E^2\left(z\right) = \Omega_{m_0} \left(1+z\right)^3 + \Omega_{r_0}\left(1+z\right)^4 + \frac{1 - \Omega_{m_0} - \Omega_{r_0}}{(b_2 + 1)e^{-b_2} - 1} \left[\left(1 + b_2 E(z)\right) \text{Exp}\left[-b_2 E(z)\right] - 1\right]\,,
\end{equation}
which reduces to $\Lambda$CDM as $b_2 \rightarrow +\infty$. We perform our analysis for $1/b_2$ so that this limit becomes $1/b_2 \rightarrow 0^{+}$, which also makes the numerical approach more stable.

\begin{figure}[h]
\begin{minipage}{0.5\textwidth}
        \includegraphics[width = 1\textwidth]{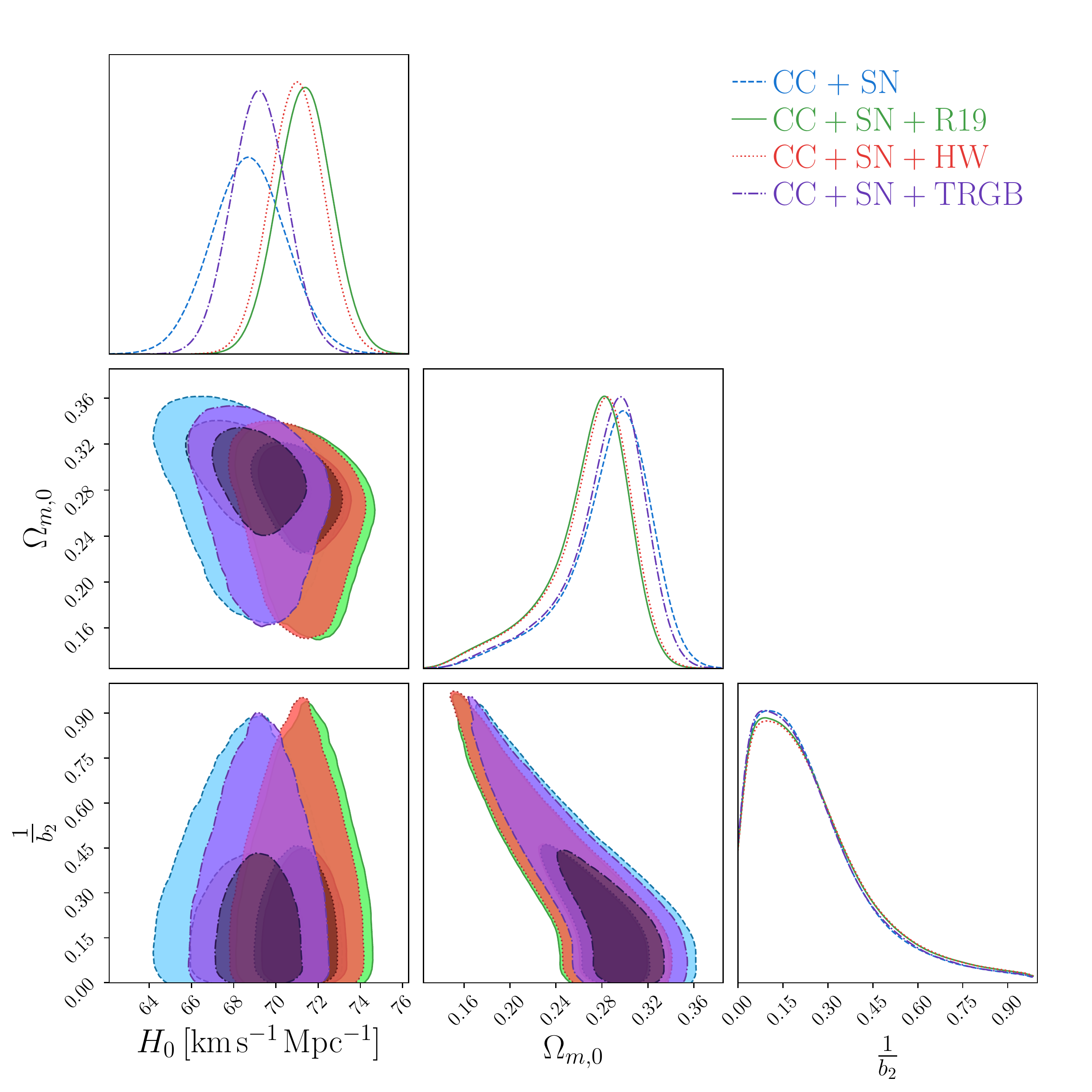}
\end{minipage}
\begin{minipage}{0.5\textwidth}
        \includegraphics[width = 1\textwidth]{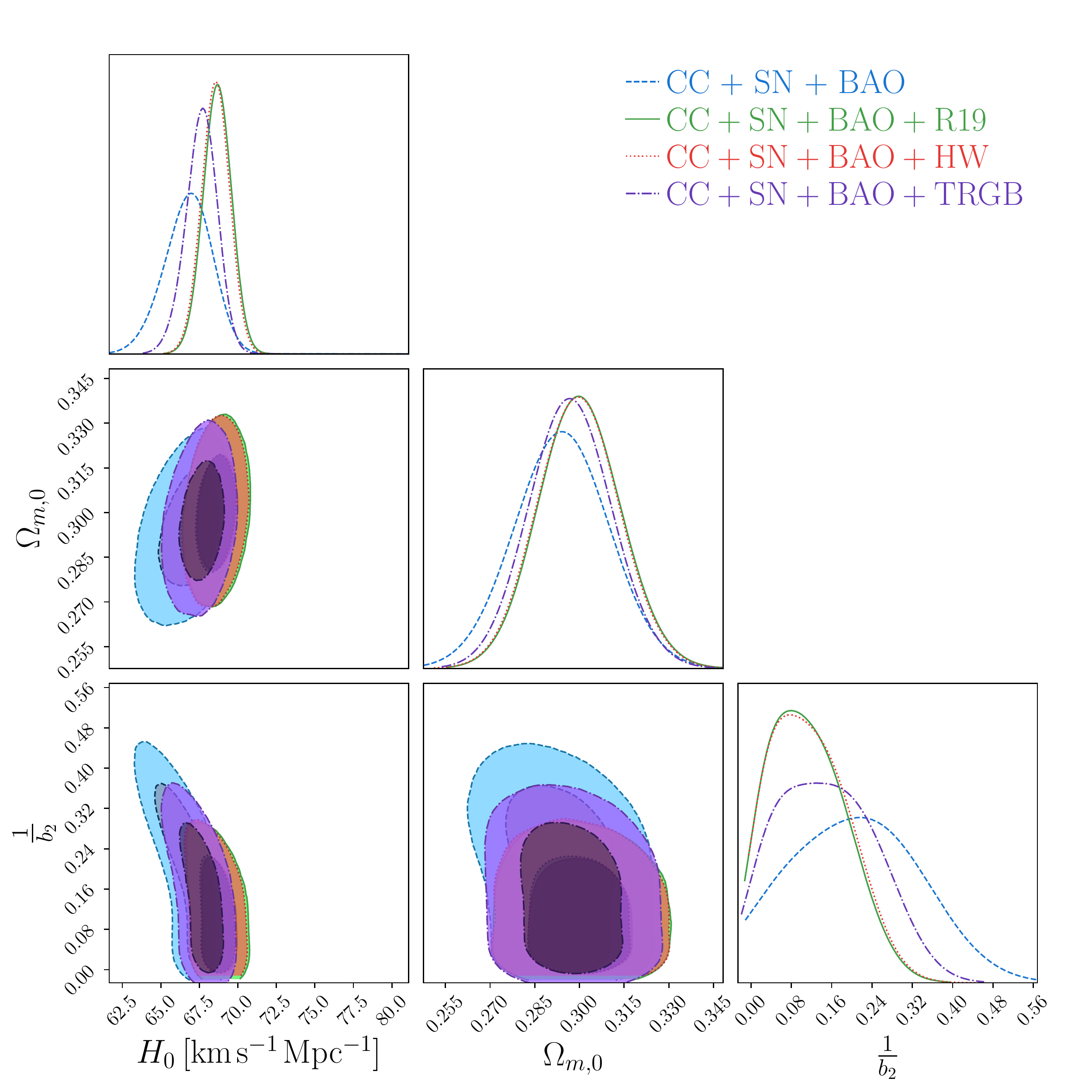}
\end{minipage}
  \caption{\textit{Left:} C.L and posteriors for the $f_2$CDM (Linder) model (\ref{eq:f2}) using CC+SN data along the $H_0$ priors: R19 (green color), HW (red color) and TRGB (purple color), respectively. \textit{Right:} C.L and posteriors for the Linder model (\ref{eq:f2}) using CC+SN+BAO data along the same priors denoted.}
    \label{fig:f2CCSN}
\end{figure}

\begin{table}
\tiny
    \centering
    \caption{Results for the $f_2$CDM (Linder) model (\ref{eq:f2}). First column: data sets used to constrain the model, including the $H_0$ priors. Second column: $H_0$ values derived from the analysis. Third column: Constrained $\Omega_{m,0}$. Fourth column: Best fit $1/b_2$ values. Fifth column: Nuisance parameters, $M$. Sixth column: $\chi^{2}_{\text{min}}$. From seventh up to tenth column: AIC and BIC with their respective differences with $\Lambda$CDM within the corresponding data set and prior combination (as shown in Appendix~\ref{sec:app}).}
    \label{tab:Linder_model}
    \begin{tabular}{>{\centering}m{0.19\textwidth}>{\centering}m{0.1\textwidth}>{\centering}m{0.1\textwidth}>{\centering}m{0.085\textwidth}>{\centering}m{0.105\textwidth}>{\centering}m{0.04\textwidth}>{\centering}m{0.04\textwidth}>{\centering}m{0.04\textwidth}>{\centering}m{0.02\textwidth}>{\centering\arraybackslash}m{0.02\textwidth}}
        \hline
		\rule{0pt}{1.2 \baselineskip}\rule{0pt}{1.2 \baselineskip}Data Sets & $H_0$[\si{\km/ \s / \mega \parsec}] & $\Omega_{m,0}$ & $\frac{1}{b_{2}}$ & M & $\chi^2_\mathrm{min}$& AIC & BIC & $\Delta$AIC & $\Delta$BIC \\ [4pt]
		\hline
		\rule{0pt}{1.2 \baselineskip}CC + SN & $68.7^{+1.8}_{-1.7}$ & $0.298^{+0.031}_{-0.035}$ & $0.101^{+0.227}_{-0.098}$ & $-19.43^{+0.57}_{-0.47}$ & 1041.49 & 1049.49 & 1069.43 & 2.00 & 6.98 \\[4pt] 
		$ \mathrm{CC + SN + R19}$ & $71.4\pm 1.3$ & $0.283^{+0.027}_{-0.036}$ & $0.088^{+0.252}_{-0.086}$ & $-19.28^{+0.50}_{-0.51}$ & 1046.32 & 1054.32 & 1074.25 & 2.00 & 6.99 \\ [4pt]
		$\mathrm{CC+ SN + HW}$ & $71.0^{+1.3}_{-1.2}$ & $0.285^{+0.027}_{-0.036}$ & $0.096^{+0.245}_{-0.093}$ & $-19.37^{+0.45}_{-0.34}$ & 1044.99 & 1052.99 & 1072.93 & 2.00 & 6.99 \\ [4pt]
		$\mathrm{CC + SN + TRGB}$ & $69.2 \pm 1.3$ & $0.296^{+0.028}_{-0.035}$ & $0.088^{+0.239}_{-0.085}$ & $-19.36^{+0.36}_{-0.37}$ & 1041.69 & 1049.69 & 1069.62 & 2.00 & 6.99\\ [4pt]
		\hline
		\rule{0pt}{1.2 \baselineskip}CC + SN + BAO & $66.9^{+1.5}_{-1.6}$ & $0.294\pm 0.016$ & $0.22^{+0.12}_{-0.15}$ & $-19.38^{+0.22}_{-0.35}$ & 1056.52 & 1064.62& 1084.52 & 1.06 & 6.06 \\ [4pt]
		$\mathrm{CC + SN + BAO + R19}$ & $68.71^{+0.88}_{-0.96}$ & $0.300\pm 0.014$ & $0.079^{+0.098}_{-0.064}$ & $-19.35^{+0.19}_{-0.24}$ & 1068.31 & 1076.31 & 1096.31 & 2.00 & 7.00\\ [4pt]
		$\mathrm{CC+ SN + BAO + HW}$ & $68.58^{+0.89}_{-0.92}$ & $0.300^{+0.013}_{-0.014}$ & $0.076^{+0.105}_{-0.060}$ & $-19.389^{+0.045}_{-0.047}$ &  1066.03 & 1074.03 & 1094.03 & 2.00 & 7.00\\ [4pt]
		$\mathrm{CC + SN + BAO + TRGB}$ & $67.7\pm 1.0$ & $0.297\pm 0.014$ & $0.128^{+0.111}_{-0.099}$ & $-19.46^{+0.37}_{-0.26}$ & 1058.47 & 1066.47 & 1086.47& 1.90 & 6.90\\ [4pt]
		\hline
    \end{tabular}
\end{table}

For this case, we show the posteriors and confidence regions of all the output MCMC runs in Fig.~\ref{fig:f2CCSN}. Some similar patterns to the $f_1$CDM model emerge such as priors on $H_0$ pushing the value of the Hubble constant from the MCMC runs to higher values, and some anti-correlation between $\Omega_{\rm m,0}$ and the model parameter $b_i$. However, this is a different model and has different properties. By design, this naturally leads to an accelerating late time Universe \cite{Linder:2010py}, which we do confirm in this case and, in fact we get slightly higher values for the expansion rate when compared with the $f_1$CDM power-law model.

In Table.~\ref{tab:Linder_model}, we show the precision outputs for each of the MCMC runs together with their settings for priors on $H_0$. While slightly higher, we again find that the largest $H_0$ is achieved for the CC+SN data set with an R19 prior ($H_0=71.4\pm^1.3 \si{\km \s^{-1} {\mega \parsec}^{-1}}$) which is an expected feature as in $f_1$CDM . In tandem, we then find the lowest value of $\Omega_{\rm m,0}$ for this MCMC run in this data set. Analogously, the CC+SN+BAO MCMC run with no prior ($H_0=66.9^{+1.5}_{-1.6} \si{\km \s^{-1} {\mega \parsec}^{-1}}$) gives the least expanding cosmology. Given how close the R19 and HW priors are in value, in all the MCMC runs, they always gave comparatively similar results for the ensuing cosmological parameters.

For numerical stability we take $1/b_2$ to be the active parameter in our MCMC runs, which means that the $\Lambda$CDM limit would be represented by $1/b_2 \rightarrow 0^{+}$. Dissimilar to $f_1$CDM, most models fall within a 2$\sigma$ rather than 1$\sigma$ making them slightly further away from favouring $\Lambda$CDM directly. Another curious feature of the results is that despite each of the $\Delta$AIC and $\Delta$BIC being calculated for each corresponding data set and prior combination for $\Lambda$CDM, it turns out that most $\Delta$AIC and $\Delta$BIC values tend to about 2.0 and 6.99 respectively, with the closest to $\Lambda$CDM being the CC+SN+BAO with no prior setting. Given that BAO depends on the early Universe to some extend, this is not an unexpected result for the model.

While still consistent with the literature in Refs.~\cite{Nesseris:2013jea,2018ApJ...855...89X,Benetti:2020hxp,Wang:2020zfv}, the various priors do add a lot more detail to the impact of the priors on $H_0$ and their effect on the MCMC runs and the resulting model parameter values. For instance, the priors by and large reduce the uncertainties in the resulting parameters which is interesting for purposes of comparison with $\Lambda$CDM.

\subsection{\texorpdfstring{$f_3(T)$}{}CDM Model}

Motivated by works in $f(\lc{R})$ gravity~\cite{Linder:2009jz}, Ref.~\cite{Nesseris:2013jea} proposes the following variant of the Linder model
\begin{equation}\label{eq:f3}
    \mathcal{F}_3 (T) = \alpha_3 T_0\left(1 - \text{Exp}\left[-b_3 T/T_0\right]\right)\,,
\end{equation}
where $\alpha_3$ and $b_3$ are constants, which can be related through
\begin{equation}
    \alpha_3 = \frac{1 - \Omega_{M_0}}{\left(1 + 2b_3\right) e^{-b_3} - 1}\,,
\end{equation}
where the Friedmann equation~\eqref{eq:Friedmann_1} was evaluated at current times. Thus, the Friedmann equation can be written as
\begin{equation}
    E^2\left(z\right) = \Omega_{m_0} \left(1+z\right)^3 + \Omega_{r_0}\left(1+z\right)^4 + \frac{1 - \Omega_{m_0} - \Omega_{r_0}}{(1 + 2b_3 )e^{-b_3} - 1} \left[\left(1 + 2b_3 E^2 (z)\right)\text{Exp}\left[-b_3 E^2 (z)\right] - 1\right]\,,
\end{equation}
which tends to $\Lambda$CDM as $b_3 \rightarrow +\infty$ similar to $f_2$CDM. As in $f_2$CDM, we again perform our analysis for $1/b_3$ so that the analysis is more stable. In this case, the $\Lambda$CDM limits comes as $1/b_3 \rightarrow 0^{+}$.

\begin{figure}[h]
\begin{minipage}{0.5\textwidth}
     \includegraphics[width = 1\textwidth]{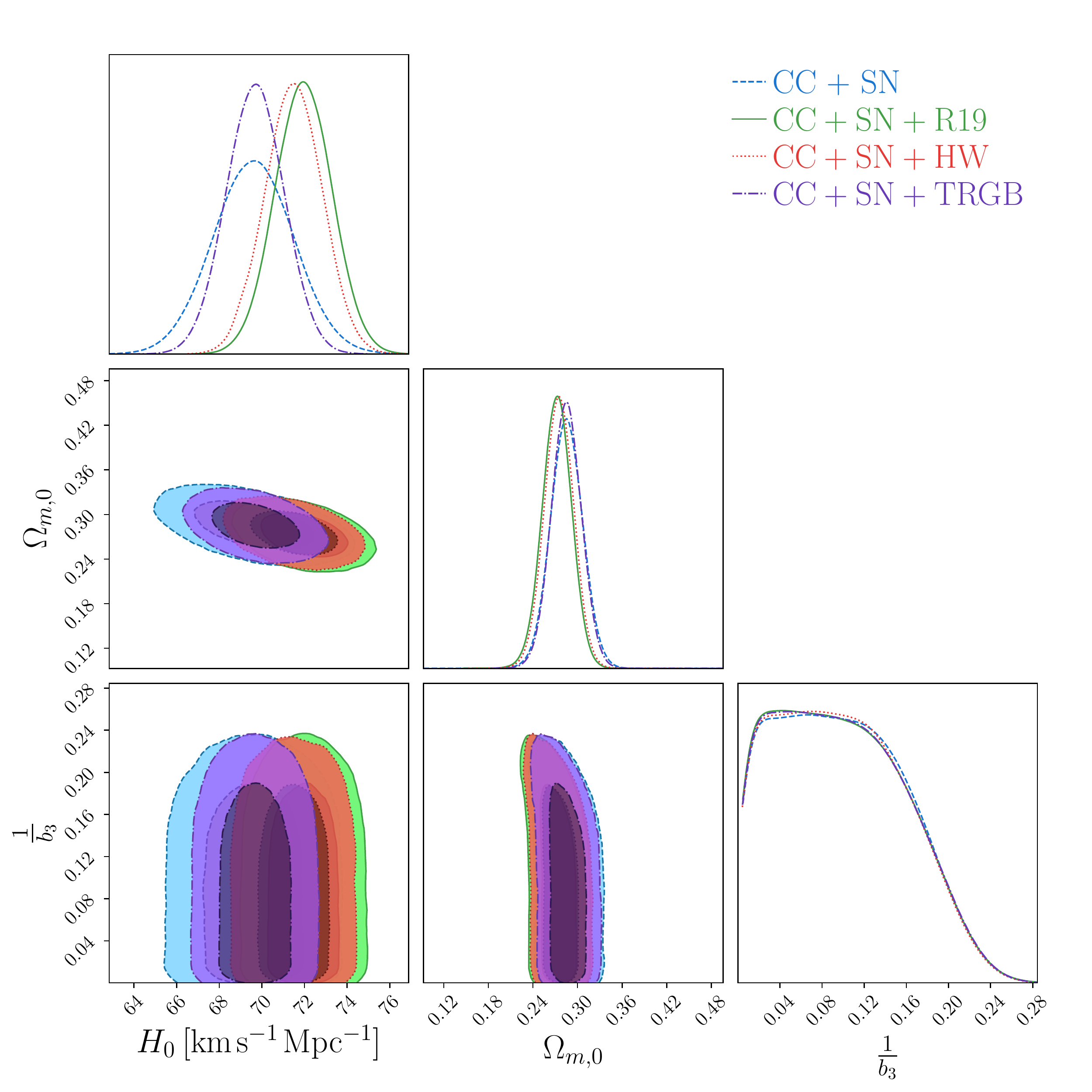}   
\end{minipage}
\begin{minipage}{0.5\textwidth}
     \includegraphics[width = 1\textwidth]{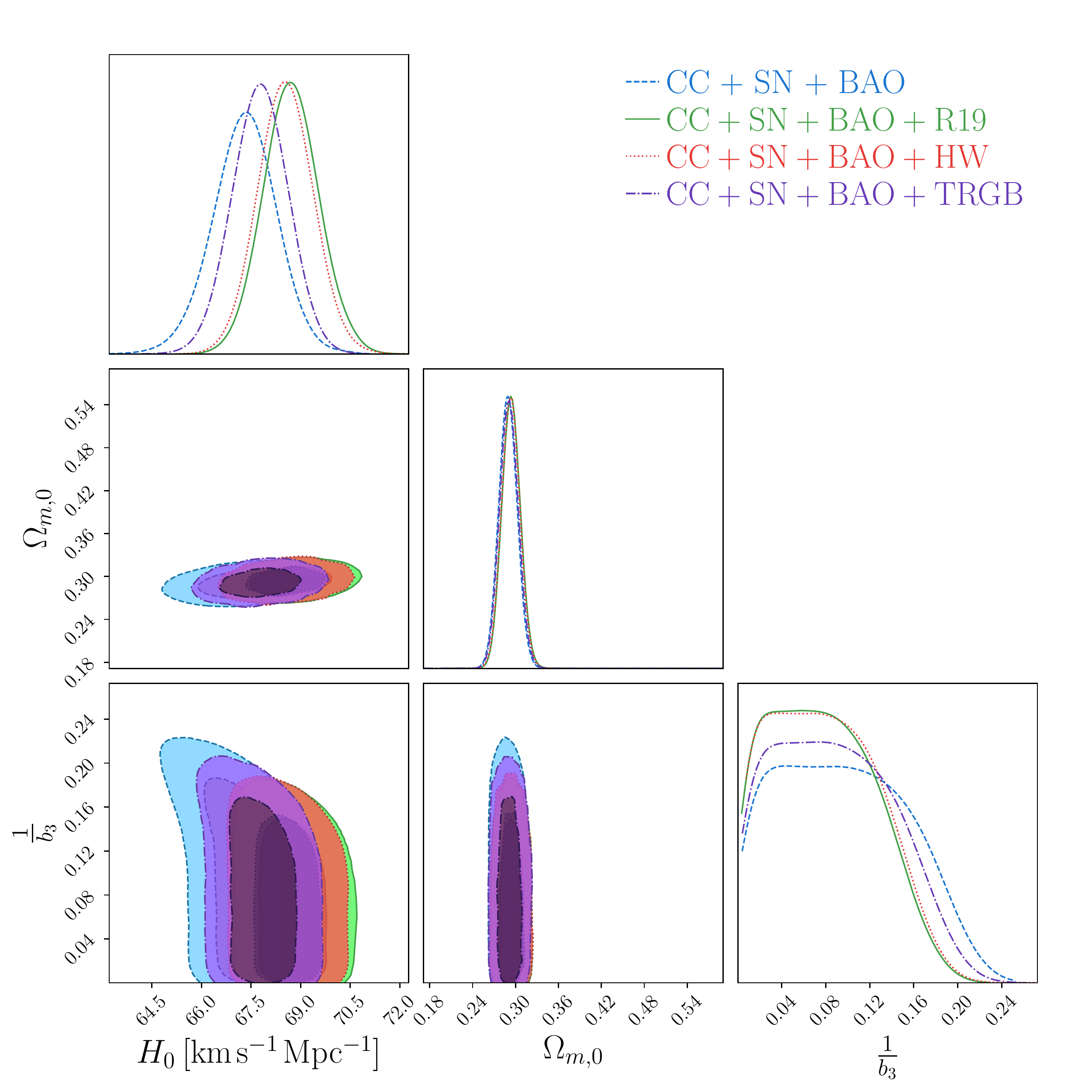}   
\end{minipage}
    \caption{\textit{Left:} C.L and posteriors for the $f_3$CDM model (\ref{eq:f3}) using CC+SN data along the $H_0$ priors: R19 (green color), HW (red color) and TRGB (purple color), respectively. \textit{Right:} C.L and posteriors for the model (\ref{eq:f3}) using CC+SN+BAO data along the same priors denoted.}
    \label{fig:f3CCSN}
\end{figure}

\begin{table}
\tiny
    \centering
    \caption{Results for the $f_3$CDM model (\ref{eq:f3}). First column: data sets used to constrain the model, including the $H_0$ priors. Second column: $H_0$ values derived from the analysis. Third column: Constrained $\Omega_{m,0}$. Fourth column: Best fit $1/b_3$ values. Fifth column: Nuisance parameter, $M$. Sixth column: $\chi^{2}_{\text{min}}$. From seventh up to tenth column: AIC and BIC with their respective differences with $\Lambda$CDM within the corresponding data set and prior combination (as shown in Appendix~\ref{sec:app}).}
    \label{tab:Linder_model2}
    \begin{tabular}{>{\centering}m{0.19\textwidth}>{\centering}m{0.09\textwidth}>{\centering}m{0.1\textwidth}>{\centering}m{0.085\textwidth}>{\centering}m{0.115\textwidth}>{\centering}m{0.04\textwidth}>{\centering}m{0.04\textwidth}>{\centering}m{0.04\textwidth}>{\centering}m{0.02\textwidth}>{\centering\arraybackslash}m{0.02\textwidth}}
        \hline
		\rule{0pt}{1.2 \baselineskip}\rule{0pt}{1.2 \baselineskip}Data Sets & $H_0$[\si{\km/ \s / \mega \parsec}] & $\Omega_{m,0}$ & $\frac{1}{b_{3}}$ & $M$ & $\chi^2_\mathrm{min}$& AIC & BIC & $\Delta$AIC & $\Delta$BIC \\ [4pt]
		\hline
		\rule{0pt}{1.2 \baselineskip}CC + SN & $69.6^{+1.8}_{-1.9}$ & $0.286^{+0.021}_{-0.023}$ & $0.067^{+0.078}_{-0.054}$  & $-19.367^{+0.053}_{-0.056}$ & 1045.04 & 1053.04 & 1072.97 & 5.55 & 10.53 \\[4pt] 
		$ \mathrm{CC + SN + R19}$ & $72.0^{+1.3}_{-1.4}$ & $0.273\pm 0.020$ & $0.042^{+0.099}_{-0.032}$  & $-19.302^{+0.037}_{-0.039}$ & 1048.16 & 1056.16 & 1076.10 & 3.84 & 8.82 \\ [4pt]
		$\mathrm{CC+ SN + HW}$ & $71.5\pm 1.4$ & $0.275^{+0.019}_{-0.020}$ & $0.070^{+0.072}_{-0.058}$ & $-19.317^{+0.039}_{-0.038}$ & 1047.06 & 1055.07 & 1075.01 & 4.08 & 9.06  \\ [4pt]
		$\mathrm{CC + SN + TRGB}$ & $69.7^{+1.3}_{-1.4}$& $0.285^{+0.020}_{-0.021}$ & $0.048^{+0.094}_{-0.037}$ & $-19.366\pm 0.040$ & 1045.04 & 1053.04 & 1072.98 & 5.36 & 10.34 \\ [4pt]
		\hline
		\rule{0pt}{1.2 \baselineskip}CC + SN + BAO & $67.35^{+0.94}_{-0.97}$ & $0.289\pm 0.013$ & $0.043^{+0.101}_{-0.026}$ & $-19.441^{+0.032}_{-0.031}$ & 1060.55 & 1068.55 & 1088.55 & 5.09 & 10.09 \\ [4pt]
		$\mathrm{CC + SN + BAO + R19}$ & $68.70^{+0.84}_{-0.85}$ & $0.293^{+0.013}_{-0.012}$ & $0.059^{+0.056}_{-0.047}$ & $-19.397^{+0.029}_{-0.028}$ & 1071.71 & 1079.71 & 1099.71 & 5.41 & 10.41\\ [4pt]
		$\mathrm{CC+ SN + BAO + HW}$ & $68.52^{+0.85}_{-0.82}$ & $0.295^{+0.011}_{-0.014}$ & $0.034^{+0.089}_{-0.024}$ & $-19.401^{+0.028}_{-0.029}$ & 1069.03 & 1077.03 & 1097.03 & 4.99 & 9.10\\ [4pt]
		$\mathrm{CC + SN + BAO + TRGB}$ & $67.79\pm 0.85$ & $0.292^{+0.012}_{-0.014}$ & $0.074^{+0.057}_{-0.059}$ & $-19.425^{+0.027}_{-0.030}$ & 1061.78 & 1069.78 & 1089.78 & 5.21 & 10.21\\ [4pt]
		\hline
    \end{tabular}
\end{table}

The posterior and confidence regions for the $f_3$CDM model are shown in Fig.~\ref{fig:f3CCSN}. This is an interesting model since in removing the square root in the exponential index there is a clear impact on the confidence regions on $\Omega_{\rm m,0}$ which now has tighter confidence regions, while $H_0$ is largely left unaffected. Saying that, we again see margin increases in the value of $H_0$ for some data set and prior settings.

The precision values shown in Table.~\ref{tab:Linder_model2} make clearer the stricter confidence levels in the density parameter values which certainly helps understand the predictions of the model better. As in the previous models, we again see the impact of priors producing a shift of the value of $H_0$ to higher values, as would be expected. Also, we observe a mildly higher maximum value of the Hubble constant ($H_0=72.0^{+1.3}_{-1.4} \si{\km \s^{-1} {\mega \parsec}^{-1}}$) with almost Gaussian errors, while the lowest value is also slightly increased with this model ($H_0=67.35^{+0.94}_{-0.97} \si{\km \s^{-1} {\mega \parsec}^{-1}}$).

Now, along the same lines of reasoning as in the $f_2$CDM model, we take $1/b_3$ as our parameter in the MCMC runs for the numerical stability of this model. While the mean values are closer to their $\Lambda$CDM limits, the confidence regions still point to a largely 2$\sigma$ distance between the $f_3$CDM and $\Lambda$CDM models. On the other hand, the $\Delta$AIC and $\Delta$BIC classifiers now give a range of values for the different MCMC settings. However, in this case, the runs favour the CC+SN MCMC run with an R19 prior.

The results presented in Fig.~\ref{fig:f3CCSN} and Table.~\ref{tab:Linder_model2} are consistent with those reported in Refs.~\cite{Nesseris:2013jea,2018ApJ...855...89X,Benetti:2020hxp,Wang:2020zfv}. However, the addition of other priors shows how the value of $H_0$ can also be smaller.

\subsection{\texorpdfstring{$f_4(T)$}{}CDM Model}
The logarithmic model, proposed in Ref.~\cite{Bamba:2010wb}, is described by
\begin{equation}\label{eq:f4}
    \mathcal{F}_4 (T) = \alpha_4 T_0 \sqrt{\frac{T}{b_4 T_0}} \log\left[\frac{b_4 T_0}{T}\right]\,,
\end{equation}
where $\alpha_4$ and $b_4$ are constants. Evaluating the Friedmann equation at current times gives
\begin{equation}
    \alpha_4 = -\frac{\left(1-\Omega_{M_0}\right)\sqrt{b_4}}{2}\,,
\end{equation}
which reduces the Friedmann equation~\eqref{eq:Friedmann_1} to the relatively simple form
\begin{equation}
    E^2\left(z\right) = \Omega_{m_0} \left(1+z\right)^3 + \Omega_{r_0}\left(1+z\right)^4 + \left(1 - \Omega_{m_0} - \Omega_{r_0} \right) E(z)\,,
\end{equation}
which interestingly does not feature $b_4$, meaning that background data cannot constrain this parameter. Another important point to highlight is that no choice of parameter values can reproduce $\Lambda$CDM for $f_4$CDM. On the other hand, the background behavior of this model does coincide with that of a spatially 
at self-accelerating branch of the DGP braneworld model \cite{Dvali:2000hr,Deffayet:2000uy}. Models of this kind are intriguing because they cannot feature confirmation bias with $\Lambda$CDM.


\begin{figure}[h]
\begin{minipage}{0.5\textwidth}
     \includegraphics[width = 1\textwidth]{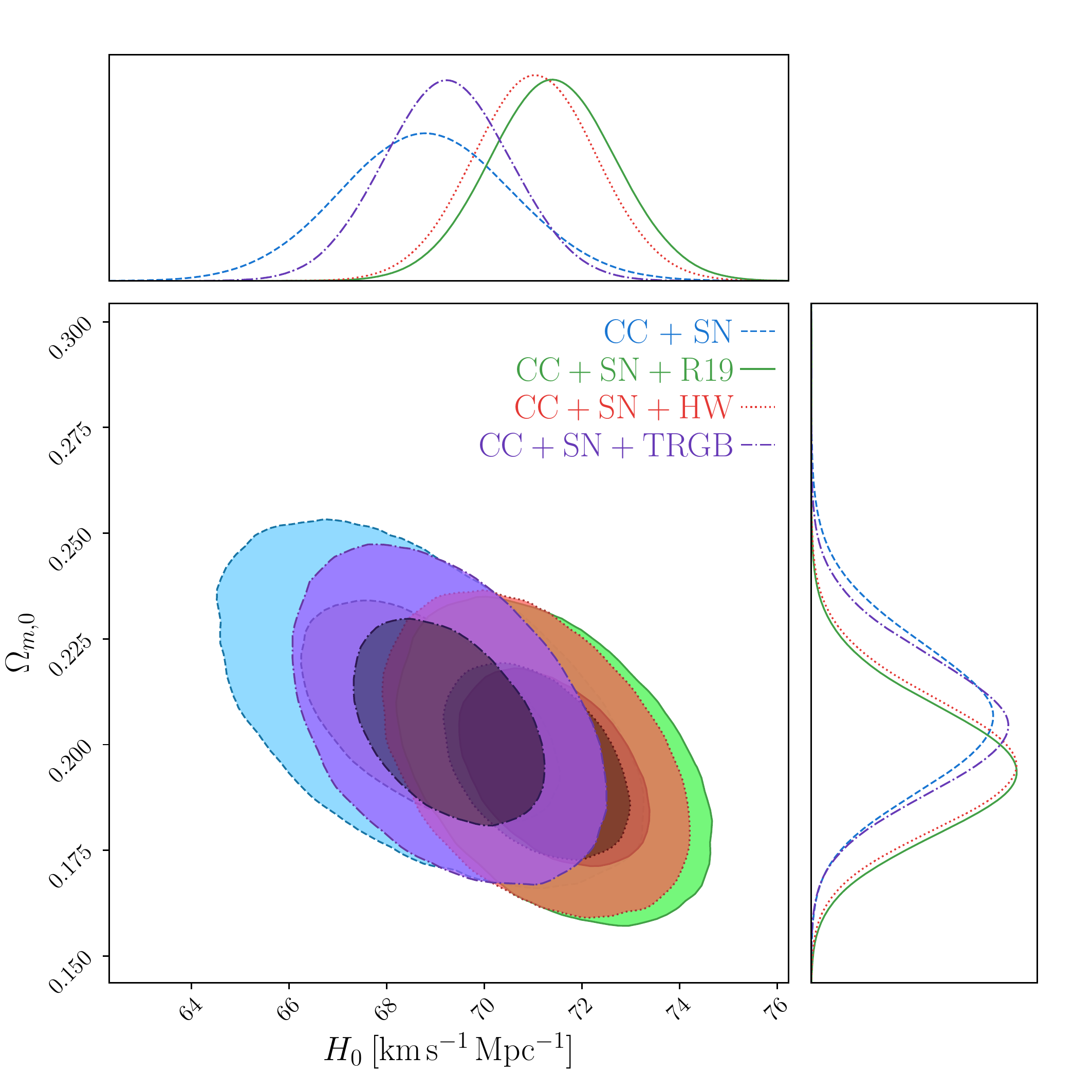}   
\end{minipage}
\begin{minipage}{0.5\textwidth}
     \includegraphics[width = 1\textwidth]{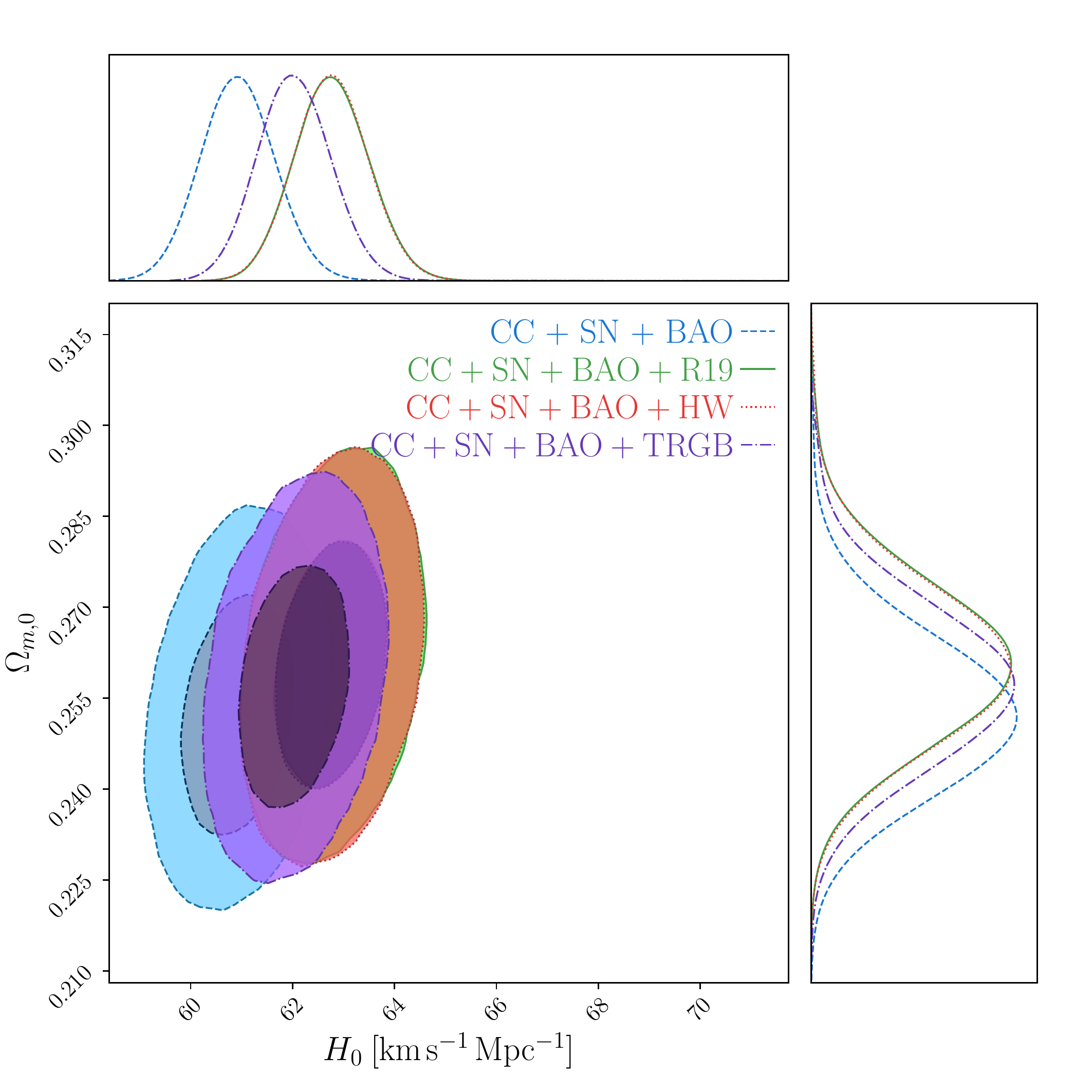}   
\end{minipage}
    \caption{\textit{Left:} C.L and posteriors for the $f_4$CDM model (\ref{eq:f4}) using CC+SN data along the $H_0$ priors: R19 (green color), HW (red color) and TRGB (purple color), respectively. \textit{Right:} C.L and posteriors for the model (\ref{eq:f4}) using CC+SN+BAO data along the same priors denoted.
    }
    \label{fig:f4CCSN}
\end{figure}

\begin{table*}[h]
\tiny
    \centering
    \caption{Results for the Logarithmic model (\ref{eq:f4}). First column: data sets used to constrain the model, including the $H_0$ priors. Second column: $H_0$ values derived from the analysis. Third column: Constrained $\Omega_{m,0}$. Fourth column: Nuisance parameter, $M$. Fifth column: $\chi^{2}_{\text{min}}$. From sixth up to ninth column: AIC and BIC with their respective differences with $\Lambda$CDM within the corresponding data set and prior combination (as shown in Appendix~\ref{sec:app}).}
    \label{tab:Log_model}
    \begin{tabular}{>{\centering}m{0.19\textwidth}>{\centering}m{0.12\textwidth}>{\centering}m{0.1\textwidth}>{\centering}m{0.12\textwidth}>{\centering}m{0.05\textwidth}>{\centering}m{0.05\textwidth}>{\centering}m{0.05\textwidth}>{\centering}m{0.05\textwidth}>{\centering\arraybackslash}m{0.05\textwidth}}
        \hline
		\rule{0pt}{1.2 \baselineskip}Data Sets & $H_0$[\si{\km/ \s / \mega \parsec}] & $\Omega_{m,0}$ & $M$ & $\chi^2_{\mathrm{min}}$ & AIC & BIC & $\Delta$AIC & $\Delta$BIC \\[4pt]
		\hline
		\rule{0pt}{1.2 \baselineskip}CC+ SN & $68.8^{+1.8}_{-1.7}$ & $0.207^{+0.018}_{-0.017}$ & $-19.371^{+0.051}_{-0.052}$ &  1043.46 & 1042.46 & 1064.41 & 1.97 & 1.97 \\ [4pt]
		$ \mathrm{CC + SN + R19}$ & $71.4\pm 1.3$ & $0.194^{+0.016}_{-0.015}$& $-19.301^{+0.038}_{-0.035}$  & 1047.98 & 1053.98 & 1068.94 & 1.67 & 1.67 \\ [4pt]
		$\mathrm{CC+ SN + HW}$ &  $71.0^{+1.3}_{-1.2}$ & $0.195^{+0.016}_{-0.015}$  & $-19.309\pm 0.036$ & 1046.70 & 1052.70 & 1067.65 & 1.70 & 1.70 \\ [4pt]
		$\mathrm{CC + SN + TRGB}$ & $69.2\pm 1.3$ & $0.205\pm 0.016$ & $-19.358^{+0.037}_{-0.040}$ & 1043.60 & 1049.60 & 1064.55 & 1.91 & 1.91\\ [4pt]
		\hline 
		\rule{0pt}{1.2 \baselineskip}CC + SN + BAO & $60.89^{+0.75}_{-0.71}$ & $0.252^{+0.014}_{-0.013}$ & $-19.611^{+0.027}_{-0.031}$ & 1078.52 & 1084.52 & 1099.52 & 21.07 & 21.07 \\ [4pt]
 		$ \mathrm{CC + SN + BAO + R19}$ & $62.77^{+0.72}_{-0.76}$ & $0.261^{+0.013}_{-0.014}$ & $-19.543^{+0.028}_{-0.029}$ & 1124.87 & 1130.87 & 1145.87 & 56.56 & 56.56 \\ [4pt]
		$\mathrm{CC+ SN + BAO +HW}$ & $62.77^{+0.70}_{-0.76}$ & $0.260^{+0.014}_{-0.013}$ & $-19.544^{+0.029}_{-0.028}$ & 1121.49 & 1127.49 & 1142.49 & 55.47 & 55.46 \\ [4pt]
		$\mathrm{CC + SN +BAO + TRGB}$ & $62.02^{+0.72}_{-0.73}$ & $0.257^{+0.013}_{-0.014}$ & $-19.567^{+0.026}_{-0.033}$ & 1097.80 & 1103.80 & 1118.80  & 39.23 & 39.23\\[4pt] 
		\hline
    \end{tabular}
\end{table*}

For this model, the posteriors and confidence regions are shown in Fig.~\ref{tab:Log_model}. This is quite an interesting model since it has no $\Lambda$CDM limit and so there is no \textit{a priori} expected values for the extra model parameter. Another crucial aspect of the $f_4$CDM model is that at background it does not contain any additional parameters in comparison to $\Lambda$CDM.

In Table.~\ref{tab:Log_model}, the precision results are presented for each of the data set and prior combinations. Here a significant difference appears depending on whether BAO data is included or not. For the cases of simply taking CC+SN data, we find reasonable values of $H_0$ with very low values of $\Omega_{\rm m,0}$ giving a maximum Hubble constant for the R19 prior ($71.4\pm 1.3 \si{\km \s^{-1} {\mega \parsec}^{-1}}$), as expected. On the other hand, when the BAO data set is included both the Hubble constant and density parameter readings turn out to be very low with a minimum $H_0 = 60.89^{+0.75}_{-0.71} \si{\km \s^{-1} {\mega \parsec}^{-1}}$. In all cases, the uncertainties are tightly constrained due to the small number of parameters.

Nothing can be said about the $b_4$ parameter from these MCMC runs since the parameter does not appear in the background equations for a flat FLRW cosmology. On the other hand, the statistical indicators tell a different story. For the CC+SN data set, irrespective of whether a prior is put on $H_0$ or not, the $\Delta$AIC and $\Delta$BIC both turn out to be low in comparison with the respective $\Lambda$CDM data set and prior combinations. However, once BAO data is, its not simply the cosmological parameters that start to veer away from their higher confidence observational regions, but also the statistical indicators. Indeed the $\Delta$AIC and $\Delta$BIC in comparison with $\Lambda$CDM both become extremely large in this scenario bringing the model into question.

In line with previous studies, \cite{Nesseris:2013jea,2018ApJ...855...89X} we find that logarithmic models do not fair well with observational data, but the addition of BAO data totally removes any possibility of the $f_4$CDM model having any significance against observational data.

\subsection{\texorpdfstring{$f_5(T)$}{}CDM Model}
Our last model under consideration is the hyperbolic tangent model given by \cite{Wu:2010av}
\begin{equation}\label{eq:f5}
    \mathcal{F}_5 (T) = \alpha_5 (-T)^{b_5} \text{Tanh}\left(\frac{T_0}{T}\right)\,,
\end{equation}
where $\alpha_5$ and $b_5$ are constants. Using Eq.~\eqref{eq:Friedmann_1} at current times results in the relation
\begin{equation}\label{eq:alphs_5_rel}
    \alpha_5 = \frac{\left(6H_0^2\right)^{1-b_5}\left(1-\Omega_{M_0}\right)}{(1 - 2b_5)\text{tanh}(1) + 2\text{Sech}^2(1)}\,.
\end{equation}
The Friedmann equation can then be written as
\begin{align}
    E^2\left(z\right) &= \Omega_{m_0} \left(1+z\right)^3 + \Omega_{r_0}\left(1+z\right)^4 +\nonumber\\
    &\frac{1-\Omega_{m_0} - \Omega_{r_0}}{(2b_5 - 1)\text{tanh}(1) - 2\text{Sech}^2(1)} E^{2(b_5-1)}(z) \left[(2b_5 - 1)E^2 (z)\text{Tanh}(E^{-2} (z)) - 2\text{Sech}^2(E^{-2} (z))\right]\,.
\end{align}
As in the $f_4$CDM model, we do not recover $\Lambda$CDM in any limit here so that $f_5$CDM does not represent a deviation from $\Lambda$CDM in so much as it is an alternative to it.

\begin{figure}[h]
\begin{minipage}{0.5\textwidth}
     \includegraphics[width = 1\textwidth]{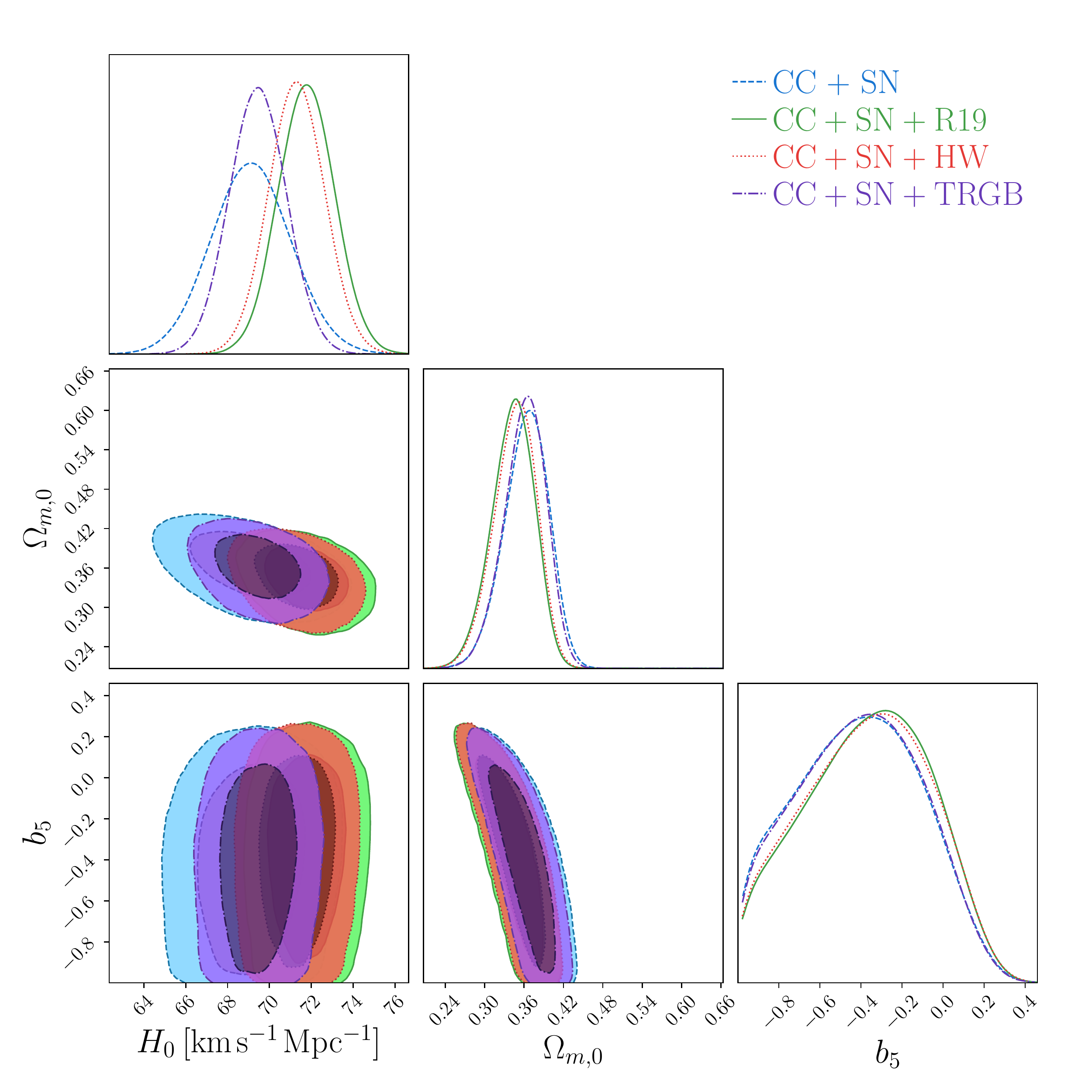}
\end{minipage}
\begin{minipage}{0.5\textwidth}
     \includegraphics[width = 1\textwidth]{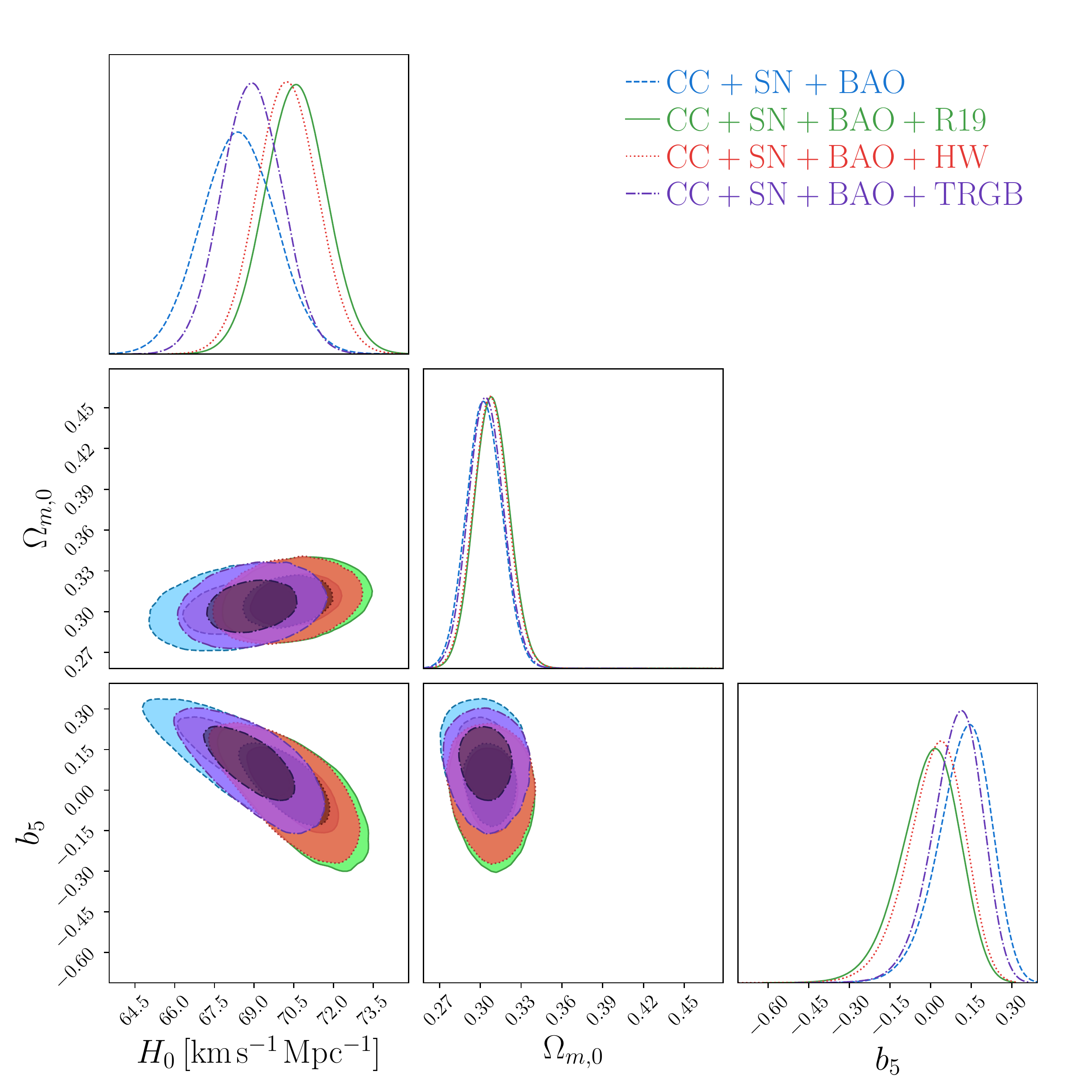}
\end{minipage}
    \caption{\textit{Left:} C.L and posteriors for the $f_5$CDM model (\ref{eq:f5}) using CC+SN data along the $H_0$ priors: R19 (green color), HW (red color) and TRGB (purple color), respectively. \textit{Right:} C.L and posteriors for the model (\ref{eq:f5}) using CC+SN+BAO data along the same priors denoted.}
    \label{fig:f5CCSN}
\end{figure}

\begin{table}
\tiny
    \centering
    \caption{Results for the hyperbolic tangent model (\ref{eq:f4}). First column: data sets used to constrain the model, including the $H_0$ priors. Second column: $H_0$ values derived from the analysis. Third column: Constrained $\Omega_{m,0}$. Fourth column: $b_5$ best fits. Fifth column: Nuisance parameter, $M$. Sixth column: $\chi^{2}_{\text{min}}$. From seventh up to tenth column: AIC and BIC with their respective differences with $\Lambda$CDM within the corresponding data set and prior combination (as shown in Appendix~\ref{sec:app}).}
    \label{tab:Hyperbolic_model}
    \begin{tabular}{>{\centering}m{0.19\textwidth}>{\centering}m{0.09\textwidth}>{\centering}m{0.1\textwidth}>{\centering}m{0.085\textwidth}>{\centering}m{0.115\textwidth}>{\centering}m{0.04\textwidth}>{\centering}m{0.04\textwidth}>{\centering}m{0.04\textwidth}>{\centering}m{0.02\textwidth}>{\centering\arraybackslash}m{0.02\textwidth}}
        \hline
		\rule{0pt}{1.2 \baselineskip}\rule{0pt}{1.2 \baselineskip}Data Sets &  $H_0$[\si{\km/ \s / \mega \parsec}] & $\Omega_{m,0}$ & $b_{5}$ & $M$ & $\chi^2_\mathrm{min}$& AIC & BIC & $\Delta$AIC & $\Delta$BIC \\ [4pt]
		\hline
		\rule{0pt}{1.2 \baselineskip}CC + SN & $69.2^{+1.9}_{-2.0}$ & $0.369^{+0.031}_{-0.036}$ & $-0.36^{+0.29}_{-0.38}$ & $-19.390^{+0.056}_{-0.058}$ & 1044.44 & 1052.44 & 1072.38 & 4.95 & 9.93 \\[4pt] 
		$ \mathrm{CC + SN + R19}$ & $71.8\pm 1.3$ & $0.349^{+0.029}_{-0.035}$ & $-0.28^{+0.28}_{-0.38}$ & $-19.314^{+0.038}_{-0.041}$ & 1048.13 & 1056.13 & 1076.06 & 3.81 & 8.79 \\ [4pt]
		$\mathrm{CC+ SN + HW}$ & $71.3\pm 1.3$ & $0.353^{+0.029}_{-0.036}$ & $-0.29^{+0.28}_{-0.38}$ & $-19.329\pm 0.040$ & 1046.94 & 1054.94 & 1074.87 & 3.94 & 8.93 \\ [4pt]
		$\mathrm{CC + SN + TRGB}$ & $69.5^{+1.3}_{-1.4}$ & $0.366^{+0.030}_{-0.034}$ & $-0.35^{+0.29}_{-0.37}$ & $-19.381^{+0.041}_{-0.042}$ & 1044.50 & 1052.50 & 1072.44 & 4.81 & 9.80\\ [4pt]
		\hline
		\rule{0pt}{1.2 \baselineskip}CC + SN + BAO & $68.4^{+1.5}_{-1.4}$ & $0.302^{+0.014}_{-0.012}$ & $0.144^{+0.087}_{-0.107}$ & $-19.400^{+0.039}_{-0.037}$ & 1062.88 & 1070.88 & 1090.88 & 7.43  & 12.43 \\ [4pt]
		$\mathrm{CC + SN + BAO + R19}$ & $70.6^{+1.1}_{-1.2}$ & $0.308\pm 0.013$ & $0.079^{+0.098}_{-0.064}$ & $-19.342^{+0.031}_{-0.033}$ & 1069.77 & 1077.77 & 1097.77 & 3.47 & 8.47\\ [4pt]
		$\mathrm{CC+ SN + BAO + HW}$ & $70.2\pm 1.1$ & $0.308^{+0.012}_{-0.013}$ & $0.039^{+0.095}_{-0.112}$ & $-19.351^{+0.032}_{-0.031}$ & 1067.90 & 1075.90 & 1095.90 & 3.88 & 8.88\\ [4pt]
		$\mathrm{CC + SN + BAO + TRGB}$ & $68.9^{+1.1}_{-1.2}$ & $0.304\pm 0.013$ & $0.115^{+0.082}_{-0.100}$ & $-19.386^{+0.032}_{-0.031}$ & 1063.34 & 1071.34 & 1091.34 & 6.78 & 11.78\\ [4pt]
		\hline
    \end{tabular}
\end{table}

In Fig.~\ref{tab:Hyperbolic_model}, the posterior and confidence regions are shown for the $f_5$CDM model. As with the $f_4$CDM model, our interest in $f_5$CDM is in its lack of a $\Lambda$CDM limit which removes any preference biases. In contrast to the $f_4$CDM model, the model parameter, namely $b_5$, does contribute at background level in the $f_5$CDM model. 

Table~\ref{tab:Hyperbolic_model} shows the precision results for the $f_5$CDM model. The situation is now drastically different to that of the $f_4$CDM model in that the spread of $H_0$ values is wider and the density parameter at current times prefers a much larger value. The maximum value of the Hubble constant which is obtained for the CC+SN data set with an R19 prior ($71.8\pm 1.3 \si{\km \s^{-1} {\mega \parsec}^{-1}}$) occurs between the $f_2$CDM and $f_3$CDM models, while the lowest value is again obtained for the CC+SN+BAO data set with no prior ($68.4^{+1.5}_{-1.4} \si{\km \s^{-1} {\mega \parsec}^{-1}}$) on $H_0$. Another issue to point out is that the uncertainties in this scenario are slightly larger than in the cases with a clear $\Lambda$CDM limit.

In this scenario the extra model parameter does play a role for the background cosmology and so has an input in the MCMC runs. By and large, the $b_5$ parameter is within the 2$\sigma$ region of being zero. However, a significant distinction occurs for the CC+SN and CC+SN+BAO data sets in that the value of $b_4$ turns out to either be negative or positive respectively. Moreover, the confidence regions are much stricter for the latter case. Another crucial difference between the $f_4$CDM and this one is that the statistical indicators are much more realistic in this scenario. Saying that, the $\Delta$AIC and $\Delta$BIC continue to feature a significant distance from $\Lambda$CDM. This calls into question whether hyperbolic tangent models are competitive scenarios for cosmology.

While adding more details with the various prior settings under consideration here, we are in agreement with the literature where this model has been investigated \cite{Nesseris:2013jea,2018ApJ...855...89X}.

\section{\label{sec:conc}Conclusion}

\begin{figure}[htbp]
    \centering
    \hbox{\hspace{-2 em}
    \includegraphics[width = \columnwidth]{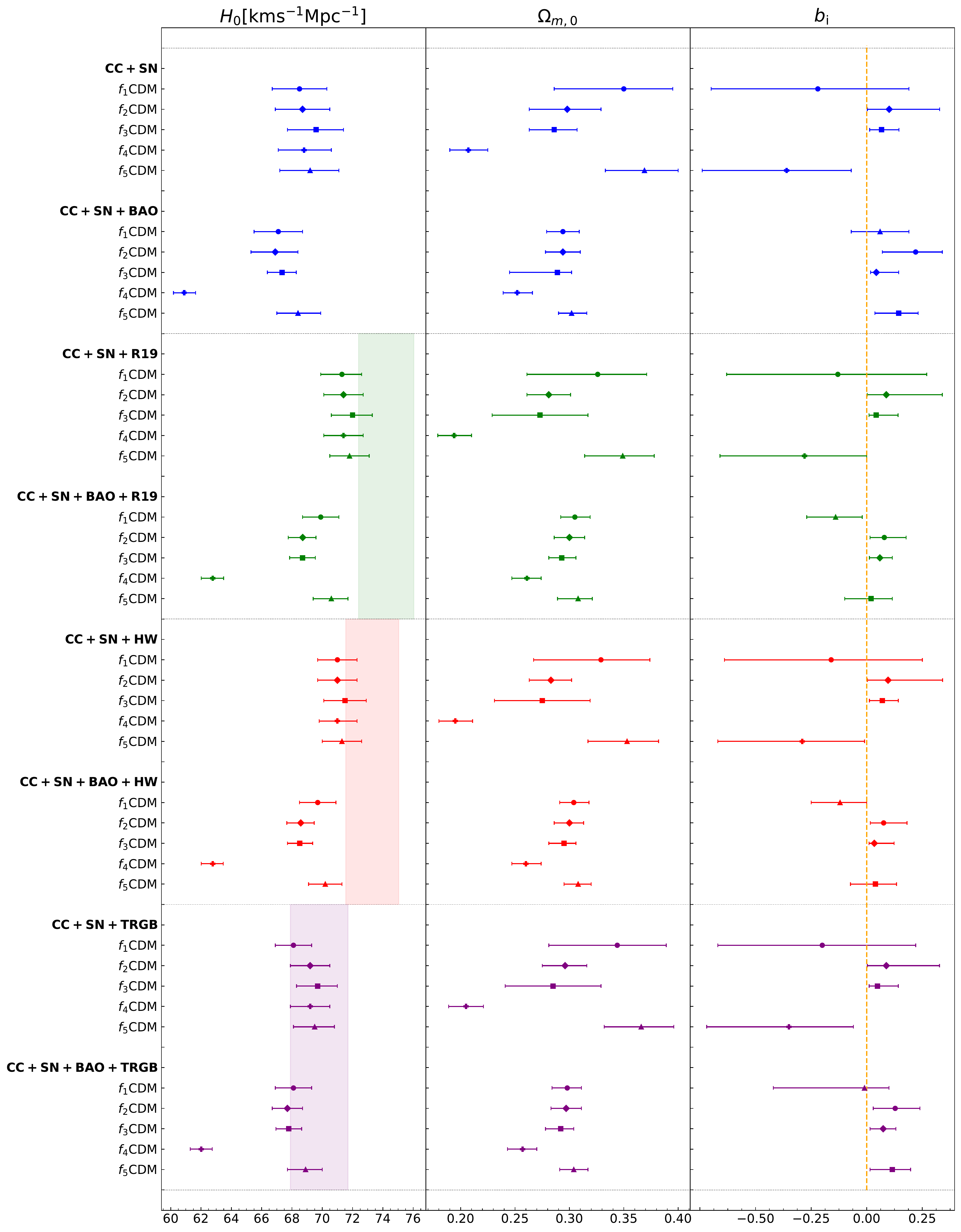}}
    \caption{Whisker plot for $H_0$, $\Omega_{m0}$ and $b_i$, respectively. The parameter $b_i$ corresponds to: $f_1$CDM as $b_1$, $f_2$CDM as $1/b_2$, $f_3$CDM as $1/b_3$, $f_4$CDM as $b_4$ and $f_5$CDM as $b_5$. The best fits are reported for CC+SN and CC+SN+BAO as: without prior (blue color), with R19 prior (green color), with HW prior (red color) and with TRGB prior (purple color). The rectangles in colors denotes the 1-$\sigma$ measured value for each project. The dashed yellow line in $b_i$ denotes the $\Lambda$CDM model.}
    \label{fig:my_label}
\end{figure}

Our study probes the impact of $H_0$ priors in $f(T)$ cosmology using five specific models together with CC, SN and BAO data sets, and 3 particularly relevant cosmology-independent priors from literature. Our interest was in assessing the way in which the cosmological parameters are altered in these various forms of $f(T)$ gravity as well as how they compare statistically with $\Lambda$CDM. To make the cross-analysis between the various models, data sets and prior choices more straightforward, we show each of the cosmological parameters against each other in the whisker plot in Fig.~\ref{fig:my_label}. Here we also show the value of each prior in shaded regions where the direct impact they have on the cosmological parameters for each model becomes much clearer.

In this work, the first three models we explore in $f_1$CDM, $f_2$CDM and $f_3$CDM are modifications of $\Lambda$CDM with a clear-cut limit in this case. We also run our MCMC analysis for $\Lambda$CDM which appears in Appendix~\ref{sec:app} which we use for our statistical analysis. These models produce higher values of $H_0$ and slightly lower values of $\Omega_{\rm m,0}$ as compared with $\Lambda$CDM at the expanse of an additional model parameter which takes the form of $b_i$ ($i=1,2,3$). This additional freedom in our cosmological model means that we can better approximate the data but produce AIC and BIC values which are higher than in $\Lambda$CDM. In our analyses, we find cosmological parameter values that are in agreement with previous works in the literature \cite{Nesseris:2013jea,2018ApJ...855...89X,Benetti:2020hxp,Wang:2020zfv} and within 1-2$\sigma$ of $\Lambda$CDM.

On the other hand, we also probe the $f_4$CDM and $f_5$CDM models which are very interesting to explore since they do not have an associated $\Lambda$CDM limit, meaning that no set of model parameter values returns an identical $\Lambda$CDM behaviour. In fact, in $f_4$CDM, the model parameter $b_4$ does not appear at background level meaning that it has the same number of parameters as $\Lambda$CDM in this regime. This also means that we have one less parameter in our MCMC analyses which means that the number of parameters in the statistical analysis will not be harshly effected by this modified form of gravity. In some case this turns out to be an advantage such as in the CC+SN runs for $f_4$CDM which have AIC and BIC values very close to $\Lambda$CDM, while for the CC+SN+BAO scenario of $\Lambda$CDM give excessively large values for these statistical quantifiers.

The work presented here probes the impact of priors on the Hubble data in the late Universe. It would be interesting to further study this effect using forecast data from future surveys at higher redshift in order to better understand how the cosmological parameters of these models may be effected in comparison to the present analysis. Furthermore, another interesting direction would be to include a perturbative analysis in which growth and gravitational wave data may be included to better probe how these priors effect those associated cosmological parameters. This will be reported elsewhere.

\begin{appendices}

\section{\texorpdfstring{$\Lambda$}{}CDM Model}\label{sec:app}

\begin{figure}[H]
    \centering
    \includegraphics[scale = 0.35]{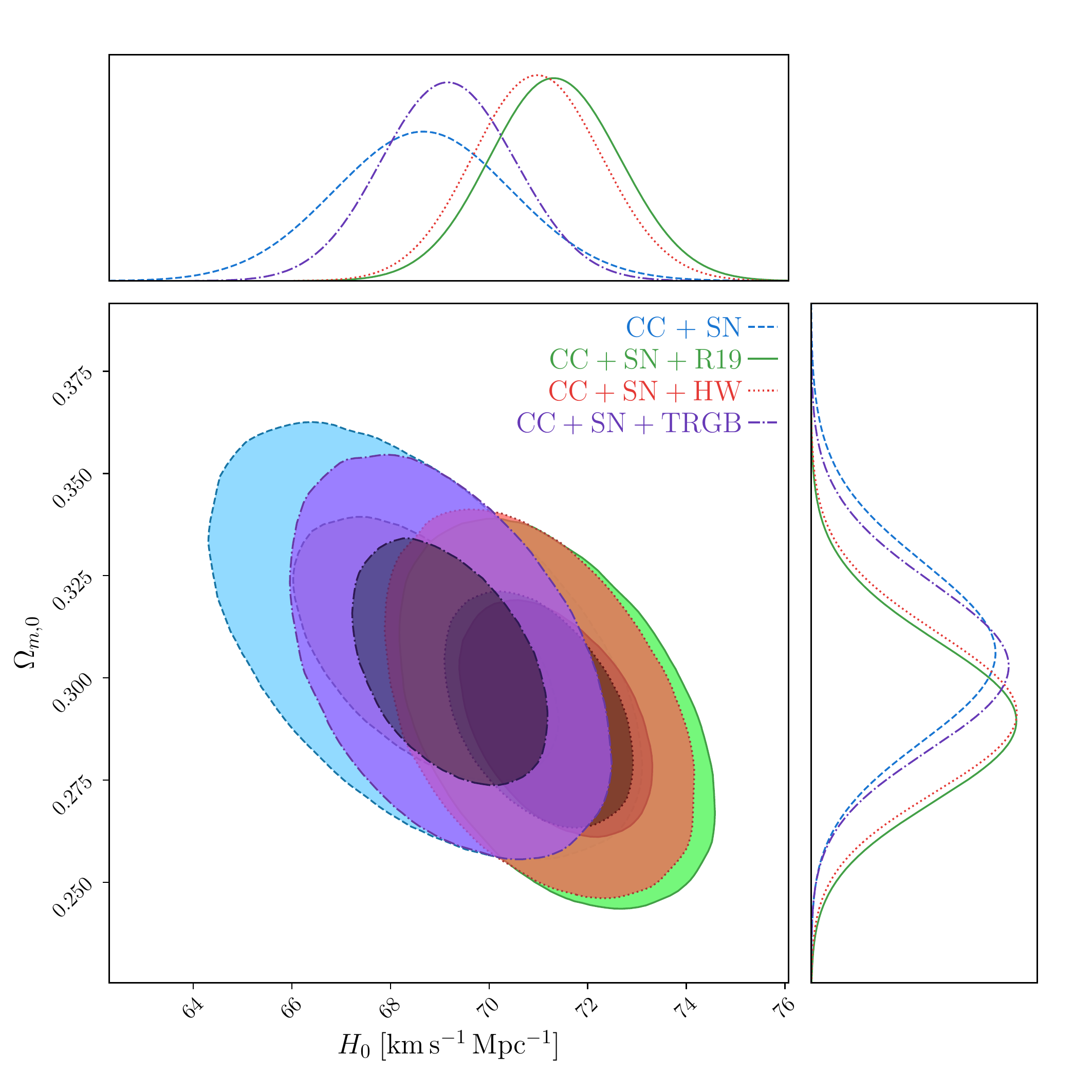}
     \includegraphics[scale = 0.35]{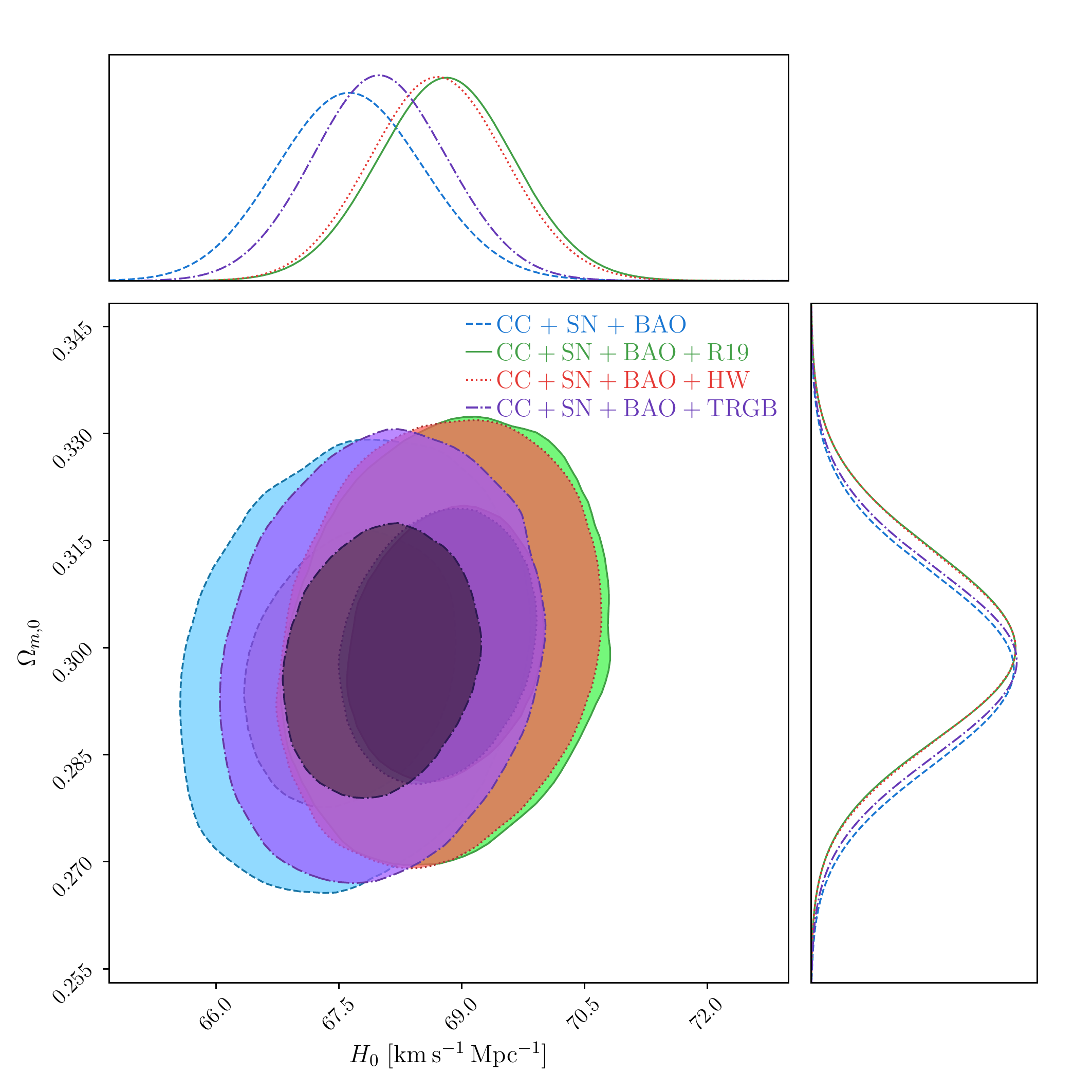}
    \caption{C.L for the $\Lambda$CDM model. \textit{Left:} C.L and posteriors for the model (\ref{eq:f4}) using CC+SN data along the $H_0$ priors: R19 (green color), HW (red color) and TRGB (purple color), respectively. \textit{Right:} C.L and posteriors for the model (\ref{eq:f4}) using CC+SN+BAO data along the same priors denoted. }
    \label{fig:f0CCSN}
\end{figure}

In all $f_i$CDM models we provide comparisons with each respective $\Lambda$CDM MCMC run. To this end, we here present the results for $\Lambda$CDM for transparency of our results. Firstly in Fig.~\ref{fig:f0CCSN}, we show the MCMC posteriors and confidence regions for the various priors for the CC+SN and CC+SN+BAO data sets. As expected, the convergence for each data set and prior combination occurs very fast giving nearly Gaussian uncertainties in every case.

\begin{table}
    \centering
    \caption{Results for the LCDM model. First column: data sets used to constrain the model, including the $H_0$ priors. Second column: $H_0$ values derived from the analysis. Third column: Constrained $\Omega_{m,0}$. Fourth column: $\chi^{2}_{\text{min}}$. From fifth-sixth column: AIC and BIC results, respectively.}
    \label{tab:LCDM_model}
    \begin{tabular}{>{\centering}m{0.25\textwidth}X>{\centering}m{0.15\textwidth}>{\centering}m{0.14\textwidth}>{\centering}m{0.1\textwidth}>{\centering}m{0.1\textwidth}>{\centering\arraybackslash}m{0.1\textwidth}}
        \hline
		\rule{0pt}{1.2 \baselineskip}Data Sets & $H_0$[\si{\km \s^{-1} {\mega \parsec}^{-1}}] & $\Omega_{m,0}$ & $\chi^2_{\mathrm{min}}$ & AIC & BIC \\[4pt]
		\hline
		\rule{0pt}{1.2 \baselineskip}CC+ SN & $68.7\pm 1.8$ & $0.306^{+0.022}_{-0.021}$ & 1041.49 & 1047.49 & 1062.44 \\ [3pt]
		$ \mathrm{CC + SN + R19}$ & $71.3^{+1.4}_{-1.3}$ & $0.290^{+0.019}_{-0.020}$ & 1046.32 & 1052.32 & 1067.27 \\ [3pt]
		$\mathrm{CC+ SN + HW}$ & $71.0\pm 1.3$ & $0.291^{+0.020}_{-0.019}$ & 1044.99 & 1050.99 & 1065.94 \\ [3pt]
		$\mathrm{CC + SN + TRGB}$ & $69.2\pm 1.4$ & $0.303^{+0.021}_{-0.020}$ & 1041.69 & 1047.69 & 1062.64 \\ [3pt]
		\hline 
		\rule{0pt}{1.2 \baselineskip}CC + SN + BAO & $67.63\pm 0.90$ & $0.297\pm 0.013$ & 1057.46 & 1063.46 & 1078.45 \\ [3pt]
		$ \mathrm{CC + SN + BAO + R19}$ & $68.81^{+0.82}_{-0.84}$ & $0.300\pm 0.013$ & 1068.30 & 1074.30 & 1089.30 \\ [3pt]
		$\mathrm{CC+ SN + BAO +HW}$ & $68.70\pm 0.83$ & $0.300^{+0.014}_{-0.013}$ & 1066.03 & 1072.03 & 1087.03 \\ [3pt]
		$\mathrm{CC + SN +BAO + TRGB}$ & $67.98^{+0.85}_{-081}$ & $0.298\pm 0.013$ & 1058.56 & 1064.56 & 1079.56 \\[3pt] 
		\hline
    \end{tabular}
\end{table}

Table~\ref{tab:LCDM_model} shows the precision results for each of the MCMC runs with the specific reference values for the AIC and BIC statistical indicators. these are the values used through the work in each of the comparison entries.

\end{appendices}


\backmatter

\bmhead*{Acknowledgments}
The authors would like to acknowledge networking support by the COST Action CA18108 and funding support from Cosmology@MALTA which is supported by the University of Malta. This research has been carried out using computational facilities procured through the European Regional Development Fund, Project No. ERDF-080 ``A supercomputing laboratory for the University of Malta''. The authors would also like to acknowledge funding from ``The Malta Council for Science and Technology'' in project IPAS-2020-007. CE-R acknowledges the Royal Astronomical Society as FRAS 10147 and by DGAPA-PAPIIT-UNAM Project IA100220.

\bibliography{references}


\begin{thebibliography}{102}
\ifx \bisbn   \undefined \def \bisbn  #1{ISBN #1}\fi
\ifx \binits  \undefined \def \binits#1{#1}\fi
\ifx \bauthor  \undefined \def \bauthor#1{#1}\fi
\ifx \batitle  \undefined \def \batitle#1{#1}\fi
\ifx \bjtitle  \undefined \def \bjtitle#1{#1}\fi
\ifx \bvolume  \undefined \def \bvolume#1{\textbf{#1}}\fi
\ifx \byear  \undefined \def \byear#1{#1}\fi
\ifx \bissue  \undefined \def \bissue#1{#1}\fi
\ifx \bfpage  \undefined \def \bfpage#1{#1}\fi
\ifx \blpage  \undefined \def \blpage #1{#1}\fi
\ifx \burl  \undefined \def \burl#1{\textsf{#1}}\fi
\ifx \doiurl  \undefined \def \doiurl#1{\url{https://doi.org/#1}}\fi
\ifx \betal  \undefined \def \betal{\textit{et al.}}\fi
\ifx \binstitute  \undefined \def \binstitute#1{#1}\fi
\ifx \binstitutionaled  \undefined \def \binstitutionaled#1{#1}\fi
\ifx \bctitle  \undefined \def \bctitle#1{#1}\fi
\ifx \beditor  \undefined \def \beditor#1{#1}\fi
\ifx \bpublisher  \undefined \def \bpublisher#1{#1}\fi
\ifx \bbtitle  \undefined \def \bbtitle#1{#1}\fi
\ifx \bedition  \undefined \def \bedition#1{#1}\fi
\ifx \bseriesno  \undefined \def \bseriesno#1{#1}\fi
\ifx \blocation  \undefined \def \blocation#1{#1}\fi
\ifx \bsertitle  \undefined \def \bsertitle#1{#1}\fi
\ifx \bsnm \undefined \def \bsnm#1{#1}\fi
\ifx \bsuffix \undefined \def \bsuffix#1{#1}\fi
\ifx \bparticle \undefined \def \bparticle#1{#1}\fi
\ifx \barticle \undefined \def \barticle#1{#1}\fi
\bibcommenthead
\ifx \bconfdate \undefined \def \bconfdate #1{#1}\fi
\ifx \botherref \undefined \def \botherref #1{#1}\fi
\ifx \url \undefined \def \url#1{\textsf{#1}}\fi
\ifx \bchapter \undefined \def \bchapter#1{#1}\fi
\ifx \bbook \undefined \def \bbook#1{#1}\fi
\ifx \bcomment \undefined \def \bcomment#1{#1}\fi
\ifx \oauthor \undefined \def \oauthor#1{#1}\fi
\ifx \citeauthoryear \undefined \def \citeauthoryear#1{#1}\fi
\ifx \endbibitem  \undefined \def \endbibitem {}\fi
\ifx \bconflocation  \undefined \def \bconflocation#1{#1}\fi
\ifx \arxivurl  \undefined \def \arxivurl#1{\textsf{#1}}\fi
\csname PreBibitemsHook\endcsname

\bibitem{Bernal:2016gxb}
\begin{barticle}
\bauthor{\bsnm{Bernal}, \binits{J.L.}},
\bauthor{\bsnm{Verde}, \binits{L.}},
\bauthor{\bsnm{Riess}, \binits{A.G.}}:
\batitle{{The trouble with $H_0$}}.
\bjtitle{JCAP}
\bvolume{10},
\bfpage{019}
(\byear{2016})
{\href{https://arxiv.org/abs/1607.05617}{{arXiv:1607.05617}}}
{[astro-ph.CO]}.
\doiurl{10.1088/1475-7516/2016/10/019}
\end{barticle}
\endbibitem

\bibitem{DiValentino:2020zio}
\begin{barticle}
\bauthor{\bsnm{Di~Valentino}, \binits{E.}}, \betal:
\batitle{{Snowmass2021 - Letter of interest cosmology intertwined II: The
  hubble constant tension}}.
\bjtitle{Astropart. Phys.}
\bvolume{131},
\bfpage{102605}
(\byear{2021})
{\href{https://arxiv.org/abs/2008.11284}{{arXiv:2008.11284}}}
{[astro-ph.CO]}.
\doiurl{10.1016/j.astropartphys.2021.102605}
\end{barticle}
\endbibitem

\bibitem{DiValentino:2021izs}
\begin{botherref}
\oauthor{\bsnm{Di~Valentino}, \binits{E.}},
\oauthor{\bsnm{Mena}, \binits{O.}},
\oauthor{\bsnm{Pan}, \binits{S.}},
\oauthor{\bsnm{Visinelli}, \binits{L.}},
\oauthor{\bsnm{Yang}, \binits{W.}},
\oauthor{\bsnm{Melchiorri}, \binits{A.}},
\oauthor{\bsnm{Mota}, \binits{D.F.}},
\oauthor{\bsnm{Riess}, \binits{A.G.}},
\oauthor{\bsnm{Silk}, \binits{J.}}:
{In the Realm of the Hubble tension $-$ a Review of Solutions}
(2021)
{\href{https://arxiv.org/abs/2103.01183}{{arXiv:2103.01183}}}
{[astro-ph.CO]}.
\doiurl{10.1088/1361-6382/ac086d}
\end{botherref}
\endbibitem

\bibitem{Planck:2018vyg}
\begin{barticle}
\bauthor{\bsnm{Aghanim}, \binits{N.}}, \betal:
\batitle{{Planck 2018 results. VI. Cosmological parameters}}.
\bjtitle{Astron. Astrophys.}
\bvolume{641},
\bfpage{6}
(\byear{2020})
{\href{https://arxiv.org/abs/1807.06209}{{arXiv:1807.06209}}}
{[astro-ph.CO]}.
\doiurl{10.1051/0004-6361/201833910}
\end{barticle}
\endbibitem

\bibitem{Baudis:2016qwx}
\begin{barticle}
\bauthor{\bsnm{Baudis}, \binits{L.}}:
\batitle{{Dark matter detection}}.
\bjtitle{J. Phys. G}
\bvolume{43}(\bissue{4}),
\bfpage{044001}
(\byear{2016}).
\doiurl{10.1088/0954-3899/43/4/044001}
\end{barticle}
\endbibitem

\bibitem{Bertone:2004pz}
\begin{barticle}
\bauthor{\bsnm{Bertone}, \binits{G.}},
\bauthor{\bsnm{Hooper}, \binits{D.}},
\bauthor{\bsnm{Silk}, \binits{J.}}:
\batitle{{Particle dark matter: Evidence, candidates and constraints}}.
\bjtitle{Phys. Rept.}
\bvolume{405},
\bfpage{279}--\blpage{390}
(\byear{2005})
{\href{https://arxiv.org/abs/hep-ph/0404175}{{arXiv:hep-ph/0404175}}}.
\doiurl{10.1016/j.physrep.2004.08.031}
\end{barticle}
\endbibitem

\bibitem{Peebles:2002gy}
\begin{barticle}
\bauthor{\bsnm{Peebles}, \binits{P.J.E.}},
\bauthor{\bsnm{Ratra}, \binits{B.}}:
\batitle{{The Cosmological Constant and Dark Energy}}.
\bjtitle{Rev. Mod. Phys.}
\bvolume{75},
\bfpage{559}--\blpage{606}
(\byear{2003})
{\href{https://arxiv.org/abs/astro-ph/0207347}{{arXiv:astro-ph/0207347}}}.
\doiurl{10.1103/RevModPhys.75.559}
\end{barticle}
\endbibitem

\bibitem{Copeland:2006wr}
\begin{barticle}
\bauthor{\bsnm{Copeland}, \binits{E.J.}},
\bauthor{\bsnm{Sami}, \binits{M.}},
\bauthor{\bsnm{Tsujikawa}, \binits{S.}}:
\batitle{{Dynamics of dark energy}}.
\bjtitle{Int. J. Mod. Phys. D}
\bvolume{15},
\bfpage{1753}--\blpage{1936}
(\byear{2006})
{\href{https://arxiv.org/abs/hep-th/0603057}{{arXiv:hep-th/0603057}}}.
\doiurl{10.1142/S021827180600942X}
\end{barticle}
\endbibitem

\bibitem{SupernovaSearchTeam:1998fmf}
\begin{barticle}
\bauthor{\bsnm{Riess}, \binits{A.G.}}, \betal:
\batitle{{Observational evidence from supernovae for an accelerating universe
  and a cosmological constant}}.
\bjtitle{Astron. J.}
\bvolume{116},
\bfpage{1009}--\blpage{1038}
(\byear{1998})
{\href{https://arxiv.org/abs/astro-ph/9805201}{{arXiv:astro-ph/9805201}}}.
\doiurl{10.1086/300499}
\end{barticle}
\endbibitem

\bibitem{SupernovaCosmologyProject:1998vns}
\begin{barticle}
\bauthor{\bsnm{Perlmutter}, \binits{S.}}, \betal:
\batitle{{Measurements of $\Omega$ and $\Lambda$ from 42 high redshift
  supernovae}}.
\bjtitle{Astrophys. J.}
\bvolume{517},
\bfpage{565}--\blpage{586}
(\byear{1999})
{\href{https://arxiv.org/abs/astro-ph/9812133}{{arXiv:astro-ph/9812133}}}.
\doiurl{10.1086/307221}
\end{barticle}
\endbibitem

\bibitem{Gaitskell:2004gd}
\begin{barticle}
\bauthor{\bsnm{Gaitskell}, \binits{R.J.}}:
\batitle{{Direct detection of dark matter}}.
\bjtitle{Ann. Rev. Nucl. Part. Sci.}
\bvolume{54},
\bfpage{315}--\blpage{359}
(\byear{2004}).
\doiurl{10.1146/annurev.nucl.54.070103.181244}
\end{barticle}
\endbibitem

\bibitem{Weinberg:1988cp}
\begin{barticle}
\bauthor{\bsnm{Weinberg}, \binits{S.}}:
\batitle{{The Cosmological Constant Problem}}.
\bjtitle{Rev. Mod. Phys.}
\bvolume{61},
\bfpage{1}--\blpage{23}
(\byear{1989}).
\doiurl{10.1103/RevModPhys.61.1}
\end{barticle}
\endbibitem

\bibitem{Riess:2019cxk}
\begin{barticle}
\bauthor{\bsnm{Riess}, \binits{A.G.}},
\bauthor{\bsnm{Casertano}, \binits{S.}},
\bauthor{\bsnm{Yuan}, \binits{W.}},
\bauthor{\bsnm{Macri}, \binits{L.M.}},
\bauthor{\bsnm{Scolnic}, \binits{D.}}:
\batitle{{Large Magellanic Cloud Cepheid Standards Provide a 1\% Foundation for
  the Determination of the Hubble Constant and Stronger Evidence for Physics
  beyond $\Lambda$CDM}}.
\bjtitle{Astrophys. J.}
\bvolume{876}(\bissue{1}),
\bfpage{85}
(\byear{2019})
{\href{https://arxiv.org/abs/1903.07603}{{arXiv:1903.07603}}}
{[astro-ph.CO]}.
\doiurl{10.3847/1538-4357/ab1422}
\end{barticle}
\endbibitem

\bibitem{Wong:2019kwg}
\begin{botherref}
\oauthor{\bsnm{Wong}, \binits{K.C.}}, et al.:
{H0LiCOW XIII. A 2.4\% measurement of $H_{0}$ from lensed quasars: $5.3\sigma$
  tension between early and late-Universe probes}
(2019)
{\href{https://arxiv.org/abs/1907.04869}{{arXiv:1907.04869}}}
{[astro-ph.CO]}
\end{botherref}
\endbibitem

\bibitem{DES:2021wwk}
\begin{botherref}
\oauthor{\bsnm{Abbott}, \binits{T.M.C.}}, et al.:
{Dark Energy Survey Year 3 Results: Cosmological Constraints from Galaxy
  Clustering and Weak Lensing}
(2021)
{\href{https://arxiv.org/abs/2105.13549}{{arXiv:2105.13549}}}
{[astro-ph.CO]}
\end{botherref}
\endbibitem

\bibitem{Graef:2018fzu}
\begin{barticle}
\bauthor{\bsnm{Graef}, \binits{L.L.}},
\bauthor{\bsnm{Benetti}, \binits{M.}},
\bauthor{\bsnm{Alcaniz}, \binits{J.S.}}:
\batitle{{Primordial gravitational waves and the H0-tension problem}}.
\bjtitle{Phys. Rev. D}
\bvolume{99}(\bissue{4}),
\bfpage{043519}
(\byear{2019})
{\href{https://arxiv.org/abs/1809.04501}{{arXiv:1809.04501}}}
{[astro-ph.CO]}.
\doiurl{10.1103/PhysRevD.99.043519}
\end{barticle}
\endbibitem

\bibitem{Abbott:2017xzu}
\begin{barticle}
\bauthor{\bsnm{Abbott}, \binits{B.P.}}, \betal:
\batitle{{A gravitational-wave standard siren measurement of the Hubble
  constant}}.
\bjtitle{Nature}
\bvolume{551}(\bissue{7678}),
\bfpage{85}--\blpage{88}
(\byear{2017})
{\href{https://arxiv.org/abs/1710.05835}{{arXiv:1710.05835}}}
{[astro-ph.CO]}.
\doiurl{10.1038/nature24471}
\end{barticle}
\endbibitem

\bibitem{Baker:2019nia}
\begin{botherref}
\oauthor{\bsnm{Baker}, \binits{J.}}, et al.:
{The Laser Interferometer Space Antenna: Unveiling the Millihertz Gravitational
  Wave Sky}
(2019)
{\href{https://arxiv.org/abs/1907.06482}{{arXiv:1907.06482}}}
{[astro-ph.IM]}
\end{botherref}
\endbibitem

\bibitem{2017arXiv170200786A}
\begin{botherref}
\oauthor{\bsnm{{Amaro-Seoane}}, \binits{P.}},
\oauthor{\bsnm{{Audley}}, \binits{H.}}, et al.:
{Laser Interferometer Space Antenna}.
arXiv e-prints,
1702--00786
(2017)
{\href{https://arxiv.org/abs/1702.00786}{{arXiv:1702.00786}}}
{[astro-ph.IM]}
\end{botherref}
\endbibitem

\bibitem{Sotiriou:2008rp}
\begin{barticle}
\bauthor{\bsnm{Sotiriou}, \binits{T.P.}},
\bauthor{\bsnm{Faraoni}, \binits{V.}}:
\batitle{{f(R) Theories Of Gravity}}.
\bjtitle{Rev. Mod. Phys.}
\bvolume{82},
\bfpage{451}--\blpage{497}
(\byear{2010})
{\href{https://arxiv.org/abs/0805.1726}{{arXiv:0805.1726}}}
{[gr-qc]}.
\doiurl{10.1103/RevModPhys.82.451}
\end{barticle}
\endbibitem

\bibitem{Clifton:2011jh}
\begin{barticle}
\bauthor{\bsnm{Clifton}, \binits{T.}},
\bauthor{\bsnm{Ferreira}, \binits{P.G.}},
\bauthor{\bsnm{Padilla}, \binits{A.}},
\bauthor{\bsnm{Skordis}, \binits{C.}}:
\batitle{{Modified Gravity and Cosmology}}.
\bjtitle{Phys. Rept.}
\bvolume{513},
\bfpage{1}--\blpage{189}
(\byear{2012})
{\href{https://arxiv.org/abs/1106.2476}{{arXiv:1106.2476}}}
{[astro-ph.CO]}.
\doiurl{10.1016/j.physrep.2012.01.001}
\end{barticle}
\endbibitem

\bibitem{CANTATA:2021ktz}
\begin{botherref}
\oauthor{\bsnm{Saridakis}, \binits{E.N.}}, et al.:
{Modified Gravity and Cosmology: An Update by the CANTATA Network}
(2021)
{\href{https://arxiv.org/abs/2105.12582}{{arXiv:2105.12582}}}
{[gr-qc]}
\end{botherref}
\endbibitem

\bibitem{Faraoni:2008mf}
\begin{bchapter}
\bauthor{\bsnm{Faraoni}, \binits{V.}}:
\bctitle{f(r) gravity: Successes and challenges}.
In: \bbtitle{18th SIGRAV Conference}
(\byear{2008})
\end{bchapter}
\endbibitem

\bibitem{Capozziello:2011et}
\begin{barticle}
\bauthor{\bsnm{Capozziello}, \binits{S.}},
\bauthor{\bsnm{De~Laurentis}, \binits{M.}}:
\batitle{{Extended Theories of Gravity}}.
\bjtitle{Phys. Rept.}
\bvolume{509},
\bfpage{167}--\blpage{321}
(\byear{2011})
{\href{https://arxiv.org/abs/1108.6266}{{arXiv:1108.6266}}}
{[gr-qc]}.
\doiurl{10.1016/j.physrep.2011.09.003}
\end{barticle}
\endbibitem

\bibitem{misner1973gravitation}
\begin{bbook}
\bauthor{\bsnm{Misner}, \binits{C.W.}},
\bauthor{\bsnm{Thorne}, \binits{K.S.}},
\bauthor{\bsnm{Wheeler}, \binits{J.A.}}:
\bbtitle{Gravitation}.
\bsertitle{Gravitation},
vol. \bseriesno{pt. 3}.
\bpublisher{W. H. Freeman}, \blocation{???}
(\byear{1973}).
\burl{https://books.google.com.mt/books?id=w4Gigq3tY1kC}
\end{bbook}
\endbibitem

\bibitem{nakahara2003geometry}
\begin{bbook}
\bauthor{\bsnm{Nakahara}, \binits{M.}}:
\bbtitle{Geometry, Topology and Physics, Second Edition}.
\bsertitle{Graduate student series in physics}.
\bpublisher{Taylor \& Francis}, \blocation{???}
(\byear{2003}).
\burl{https://books.google.com.mt/books?id=cH-XQB0Ex5wC}
\end{bbook}
\endbibitem

\bibitem{Bahamonde:2021gfp}
\begin{botherref}
\oauthor{\bsnm{Bahamonde}, \binits{S.}},
\oauthor{\bsnm{Dialektopoulos}, \binits{K.F.}},
\oauthor{\bsnm{Escamilla-Rivera}, \binits{C.}},
\oauthor{\bsnm{Farrugia}, \binits{G.}},
\oauthor{\bsnm{Gakis}, \binits{V.}},
\oauthor{\bsnm{Hendry}, \binits{M.}},
\oauthor{\bsnm{Hohmann}, \binits{M.}},
\oauthor{\bsnm{Said}, \binits{J.L.}},
\oauthor{\bsnm{Mifsud}, \binits{J.}},
\oauthor{\bsnm{Di~Valentino}, \binits{E.}}:
{Teleparallel Gravity: From Theory to Cosmology}
(2021)
{\href{https://arxiv.org/abs/2106.13793}{{arXiv:2106.13793}}}
{[gr-qc]}
\end{botherref}
\endbibitem

\bibitem{Aldrovandi:2013wha}
\begin{bbook}
\bauthor{\bsnm{Aldrovandi}, \binits{R.}},
\bauthor{\bsnm{Pereira}, \binits{J.G.}}:
\bbtitle{Teleparallel Gravity: An Introduction}.
\bpublisher{Springer}, \blocation{???}
(\byear{2013}).
\doiurl{10.1007/978-94-007-5143-9}
\end{bbook}
\endbibitem

\bibitem{Cai:2015emx}
\begin{barticle}
\bauthor{\bsnm{Cai}, \binits{Y.-F.}},
\bauthor{\bsnm{Capozziello}, \binits{S.}},
\bauthor{\bsnm{De~Laurentis}, \binits{M.}},
\bauthor{\bsnm{Saridakis}, \binits{E.N.}}:
\batitle{{$f(T)$ teleparallel gravity and cosmology}}.
\bjtitle{Rept. Prog. Phys.}
\bvolume{79}(\bissue{10}),
\bfpage{106901}
(\byear{2016})
{\href{https://arxiv.org/abs/1511.07586}{{arXiv:1511.07586}}}
{[gr-qc]}.
\doiurl{10.1088/0034-4885/79/10/106901}
\end{barticle}
\endbibitem

\bibitem{Krssak:2018ywd}
\begin{barticle}
\bauthor{\bsnm{Krssak}, \binits{M.}},
\bauthor{\bparticle{van~den} \bsnm{Hoogen}, \binits{R.J.}},
\bauthor{\bsnm{Pereira}, \binits{J.G.}},
\bauthor{\bsnm{B\"ohmer}, \binits{C.G.}},
\bauthor{\bsnm{Coley}, \binits{A.A.}}:
\batitle{{Teleparallel theories of gravity: illuminating a fully invariant
  approach}}.
\bjtitle{Class. Quant. Grav.}
\bvolume{36}(\bissue{18}),
\bfpage{183001}
(\byear{2019})
{\href{https://arxiv.org/abs/1810.12932}{{arXiv:1810.12932}}}
{[gr-qc]}.
\doiurl{10.1088/1361-6382/ab2e1f}
\end{barticle}
\endbibitem

\bibitem{Weitzenbock1923}
\begin{bbook}
\bauthor{\bsnm{Weitzenb\"{o}ock}, \binits{R.}}:
\bbtitle{`Invariantentheorie'}.
\bpublisher{Noordhoff, Gronningen}, \blocation{???}
(\byear{1923})
\end{bbook}
\endbibitem

\bibitem{Lovelock:1971yv}
\begin{barticle}
\bauthor{\bsnm{Lovelock}, \binits{D.}}:
\batitle{{The Einstein tensor and its generalizations}}.
\bjtitle{J. Math. Phys.}
\bvolume{12},
\bfpage{498}--\blpage{501}
(\byear{1971}).
\doiurl{10.1063/1.1665613}
\end{barticle}
\endbibitem

\bibitem{Gonzalez:2015sha}
\begin{barticle}
\bauthor{\bsnm{Gonzalez}, \binits{P.A.}},
\bauthor{\bsnm{Vasquez}, \binits{Y.}}:
\batitle{{Teleparallel Equivalent of Lovelock Gravity}}.
\bjtitle{Phys. Rev. D}
\bvolume{92}(\bissue{12}),
\bfpage{124023}
(\byear{2015})
{\href{https://arxiv.org/abs/1508.01174}{{arXiv:1508.01174}}}
{[hep-th]}.
\doiurl{10.1103/PhysRevD.92.124023}
\end{barticle}
\endbibitem

\bibitem{Bahamonde:2019shr}
\begin{barticle}
\bauthor{\bsnm{Bahamonde}, \binits{S.}},
\bauthor{\bsnm{Dialektopoulos}, \binits{K.F.}},
\bauthor{\bsnm{Levi~Said}, \binits{J.}}:
\batitle{{Can Horndeski Theory be recast using Teleparallel Gravity?}}
\bjtitle{Phys. Rev. D}
\bvolume{100}(\bissue{6}),
\bfpage{064018}
(\byear{2019})
{\href{https://arxiv.org/abs/1904.10791}{{arXiv:1904.10791}}}
{[gr-qc]}.
\doiurl{10.1103/PhysRevD.100.064018}
\end{barticle}
\endbibitem

\bibitem{Ferraro:2006jd}
\begin{barticle}
\bauthor{\bsnm{Ferraro}, \binits{R.}},
\bauthor{\bsnm{Fiorini}, \binits{F.}}:
\batitle{{Modified teleparallel gravity: Inflation without inflaton}}.
\bjtitle{Phys. Rev.}
\bvolume{D75},
\bfpage{084031}
(\byear{2007})
{\href{https://arxiv.org/abs/gr-qc/0610067}{{arXiv:gr-qc/0610067}}}
{[gr-qc]}.
\doiurl{10.1103/PhysRevD.75.084031}
\end{barticle}
\endbibitem

\bibitem{Ferraro:2008ey}
\begin{barticle}
\bauthor{\bsnm{Ferraro}, \binits{R.}},
\bauthor{\bsnm{Fiorini}, \binits{F.}}:
\batitle{{On Born-Infeld Gravity in Weitzenbock spacetime}}.
\bjtitle{Phys. Rev.}
\bvolume{D78},
\bfpage{124019}
(\byear{2008})
{\href{https://arxiv.org/abs/0812.1981}{{arXiv:0812.1981}}}
{[gr-qc]}.
\doiurl{10.1103/PhysRevD.78.124019}
\end{barticle}
\endbibitem

\bibitem{Bengochea:2008gz}
\begin{barticle}
\bauthor{\bsnm{Bengochea}, \binits{G.R.}},
\bauthor{\bsnm{Ferraro}, \binits{R.}}:
\batitle{{Dark torsion as the cosmic speed-up}}.
\bjtitle{Phys. Rev.}
\bvolume{D79},
\bfpage{124019}
(\byear{2009})
{\href{https://arxiv.org/abs/0812.1205}{{arXiv:0812.1205}}}
{[astro-ph]}.
\doiurl{10.1103/PhysRevD.79.124019}
\end{barticle}
\endbibitem

\bibitem{Linder:2010py}
\begin{barticle}
\bauthor{\bsnm{Linder}, \binits{E.V.}}:
\batitle{{Einstein's Other Gravity and the Acceleration of the Universe}}.
\bjtitle{Phys. Rev.}
\bvolume{D81},
\bfpage{127301}
(\byear{2010})
{\href{https://arxiv.org/abs/1005.3039}{{arXiv:1005.3039}}}
{[astro-ph.CO]}.
\doiurl{10.1103/PhysRevD.81.127301, 10.1103/PhysRevD.82.109902}.
\bcomment{[Erratum: Phys. Rev.D82,109902(2010)]}
\end{barticle}
\endbibitem

\bibitem{Chen:2010va}
\begin{barticle}
\bauthor{\bsnm{Chen}, \binits{S.-H.}},
\bauthor{\bsnm{Dent}, \binits{J.B.}},
\bauthor{\bsnm{Dutta}, \binits{S.}},
\bauthor{\bsnm{Saridakis}, \binits{E.N.}}:
\batitle{{Cosmological perturbations in f(T) gravity}}.
\bjtitle{Phys. Rev.}
\bvolume{D83},
\bfpage{023508}
(\byear{2011})
{\href{https://arxiv.org/abs/1008.1250}{{arXiv:1008.1250}}}
{[astro-ph.CO]}.
\doiurl{10.1103/PhysRevD.83.023508}
\end{barticle}
\endbibitem

\bibitem{Bahamonde:2019zea}
\begin{barticle}
\bauthor{\bsnm{Bahamonde}, \binits{S.}},
\bauthor{\bsnm{Flathmann}, \binits{K.}},
\bauthor{\bsnm{Pfeifer}, \binits{C.}}:
\batitle{{Photon sphere and perihelion shift in weak $f(T)$ gravity}}.
\bjtitle{Phys. Rev. D}
\bvolume{100}(\bissue{8}),
\bfpage{084064}
(\byear{2019})
{\href{https://arxiv.org/abs/1907.10858}{{arXiv:1907.10858}}}
{[gr-qc]}.
\doiurl{10.1103/PhysRevD.100.084064}
\end{barticle}
\endbibitem

\bibitem{Farrugia:2016qqe}
\begin{barticle}
\bauthor{\bsnm{Farrugia}, \binits{G.}},
\bauthor{\bsnm{Levi~Said}, \binits{J.}}:
\batitle{{Stability of the flat FLRW metric in $f(T)$ gravity}}.
\bjtitle{Phys. Rev. D}
\bvolume{94}(\bissue{12}),
\bfpage{124054}
(\byear{2016})
{\href{https://arxiv.org/abs/1701.00134}{{arXiv:1701.00134}}}
{[gr-qc]}.
\doiurl{10.1103/PhysRevD.94.124054}
\end{barticle}
\endbibitem

\bibitem{Finch:2018gkh}
\begin{barticle}
\bauthor{\bsnm{Finch}, \binits{A.}},
\bauthor{\bsnm{Said}, \binits{J.L.}}:
\batitle{{Galactic Rotation Dynamics in f(T) gravity}}.
\bjtitle{Eur. Phys. J. C}
\bvolume{78}(\bissue{7}),
\bfpage{560}
(\byear{2018})
{\href{https://arxiv.org/abs/1806.09677}{{arXiv:1806.09677}}}
{[astro-ph.GA]}.
\doiurl{10.1140/epjc/s10052-018-6028-1}
\end{barticle}
\endbibitem

\bibitem{Farrugia:2016xcw}
\begin{barticle}
\bauthor{\bsnm{Farrugia}, \binits{G.}},
\bauthor{\bsnm{Levi~Said}, \binits{J.}},
\bauthor{\bsnm{Ruggiero}, \binits{M.L.}}:
\batitle{{Solar System tests in f(T) gravity}}.
\bjtitle{Phys. Rev. D}
\bvolume{93}(\bissue{10}),
\bfpage{104034}
(\byear{2016})
{\href{https://arxiv.org/abs/1605.07614}{{arXiv:1605.07614}}}
{[gr-qc]}.
\doiurl{10.1103/PhysRevD.93.104034}
\end{barticle}
\endbibitem

\bibitem{Iorio:2012cm}
\begin{barticle}
\bauthor{\bsnm{Iorio}, \binits{L.}},
\bauthor{\bsnm{Saridakis}, \binits{E.N.}}:
\batitle{{Solar system constraints on f(T) gravity}}.
\bjtitle{Mon. Not. Roy. Astron. Soc.}
\bvolume{427},
\bfpage{1555}
(\byear{2012})
{\href{https://arxiv.org/abs/1203.5781}{{arXiv:1203.5781}}}
{[gr-qc]}.
\doiurl{10.1111/j.1365-2966.2012.21995.x}
\end{barticle}
\endbibitem

\bibitem{Deng:2018ncg}
\begin{barticle}
\bauthor{\bsnm{Deng}, \binits{X.-M.}}:
\batitle{{Probing f(T) gravity with gravitational time advancement}}.
\bjtitle{Class. Quant. Grav.}
\bvolume{35}(\bissue{17}),
\bfpage{175013}
(\byear{2018}).
\doiurl{10.1088/1361-6382/aad391}
\end{barticle}
\endbibitem

\bibitem{Nesseris:2013jea}
\begin{barticle}
\bauthor{\bsnm{Nesseris}, \binits{S.}},
\bauthor{\bsnm{Basilakos}, \binits{S.}},
\bauthor{\bsnm{Saridakis}, \binits{E.N.}},
\bauthor{\bsnm{Perivolaropoulos}, \binits{L.}}:
\batitle{{Viable $f(T)$ models are practically indistinguishable from
  $\Lambda$CDM}}.
\bjtitle{Phys. Rev. D}
\bvolume{88},
\bfpage{103010}
(\byear{2013})
{\href{https://arxiv.org/abs/1308.6142}{{arXiv:1308.6142}}}
{[astro-ph.CO]}.
\doiurl{10.1103/PhysRevD.88.103010}
\end{barticle}
\endbibitem

\bibitem{Anagnostopoulos:2019miu}
\begin{barticle}
\bauthor{\bsnm{Anagnostopoulos}, \binits{F.K.}},
\bauthor{\bsnm{Basilakos}, \binits{S.}},
\bauthor{\bsnm{Saridakis}, \binits{E.N.}}:
\batitle{{Bayesian analysis of $f(T)$ gravity using $f\sigma_8$ data}}.
\bjtitle{Phys. Rev. D}
\bvolume{100}(\bissue{8}),
\bfpage{083517}
(\byear{2019})
{\href{https://arxiv.org/abs/1907.07533}{{arXiv:1907.07533}}}
{[astro-ph.CO]}.
\doiurl{10.1103/PhysRevD.100.083517}
\end{barticle}
\endbibitem

\bibitem{Nunes:2018evm}
\begin{barticle}
\bauthor{\bsnm{Nunes}, \binits{R.C.}},
\bauthor{\bsnm{Pan}, \binits{S.}},
\bauthor{\bsnm{Saridakis}, \binits{E.N.}}:
\batitle{{New observational constraints on $f(T)$ gravity through
  gravitational-wave astronomy}}.
\bjtitle{Phys. Rev. D}
\bvolume{98}(\bissue{10}),
\bfpage{104055}
(\byear{2018})
{\href{https://arxiv.org/abs/1810.03942}{{arXiv:1810.03942}}}
{[gr-qc]}.
\doiurl{10.1103/PhysRevD.98.104055}
\end{barticle}
\endbibitem

\bibitem{Benetti:2020hxp}
\begin{barticle}
\bauthor{\bsnm{Benetti}, \binits{M.}},
\bauthor{\bsnm{Capozziello}, \binits{S.}},
\bauthor{\bsnm{Lambiase}, \binits{G.}}:
\batitle{{Updating constraints on f(T) teleparallel cosmology and the
  consistency with Big Bang Nucleosynthesis}}.
\bjtitle{Mon. Not. Roy. Astron. Soc.}
\bvolume{500}(\bissue{2}),
\bfpage{1795}--\blpage{1805}
(\byear{2020})
{\href{https://arxiv.org/abs/2006.15335}{{arXiv:2006.15335}}}
{[astro-ph.CO]}.
\doiurl{10.1093/mnras/staa3368}
\end{barticle}
\endbibitem

\bibitem{Bahamonde:2015zma}
\begin{barticle}
\bauthor{\bsnm{Bahamonde}, \binits{S.}},
\bauthor{\bsnm{B\"ohmer}, \binits{C.G.}},
\bauthor{\bsnm{Wright}, \binits{M.}}:
\batitle{{Modified teleparallel theories of gravity}}.
\bjtitle{Phys. Rev. D}
\bvolume{92}(\bissue{10}),
\bfpage{104042}
(\byear{2015})
{\href{https://arxiv.org/abs/1508.05120}{{arXiv:1508.05120}}}
{[gr-qc]}.
\doiurl{10.1103/PhysRevD.92.104042}
\end{barticle}
\endbibitem

\bibitem{Bahamonde:2016grb}
\begin{barticle}
\bauthor{\bsnm{Bahamonde}, \binits{S.}},
\bauthor{\bsnm{Capozziello}, \binits{S.}}:
\batitle{{Noether Symmetry Approach in $f(T,B)$ teleparallel cosmology}}.
\bjtitle{Eur. Phys. J. C}
\bvolume{77}(\bissue{2}),
\bfpage{107}
(\byear{2017})
{\href{https://arxiv.org/abs/1612.01299}{{arXiv:1612.01299}}}
{[gr-qc]}.
\doiurl{10.1140/epjc/s10052-017-4677-0}
\end{barticle}
\endbibitem

\bibitem{Paliathanasis:2017flf}
\begin{barticle}
\bauthor{\bsnm{Paliathanasis}, \binits{A.}}:
\batitle{{de Sitter and Scaling solutions in a higher-order modified
  teleparallel theory}}.
\bjtitle{JCAP}
\bvolume{08},
\bfpage{027}
(\byear{2017})
{\href{https://arxiv.org/abs/1706.02662}{{arXiv:1706.02662}}}
{[gr-qc]}.
\doiurl{10.1088/1475-7516/2017/08/027}
\end{barticle}
\endbibitem

\bibitem{Farrugia:2018gyz}
\begin{barticle}
\bauthor{\bsnm{Farrugia}, \binits{G.}},
\bauthor{\bsnm{Levi~Said}, \binits{J.}},
\bauthor{\bsnm{Gakis}, \binits{V.}},
\bauthor{\bsnm{Saridakis}, \binits{E.N.}}:
\batitle{{Gravitational Waves in Modified Teleparallel Theories}}.
\bjtitle{Phys. Rev. D}
\bvolume{97}(\bissue{12}),
\bfpage{124064}
(\byear{2018})
{\href{https://arxiv.org/abs/1804.07365}{{arXiv:1804.07365}}}
{[gr-qc]}.
\doiurl{10.1103/PhysRevD.97.124064}
\end{barticle}
\endbibitem

\bibitem{Bahamonde:2016cul}
\begin{barticle}
\bauthor{\bsnm{Bahamonde}, \binits{S.}},
\bauthor{\bsnm{Zubair}, \binits{M.}},
\bauthor{\bsnm{Abbas}, \binits{G.}}:
\batitle{{Thermodynamics and cosmological reconstruction in $f(T,B)$ gravity}}.
\bjtitle{Phys. Dark Univ.}
\bvolume{19},
\bfpage{78}--\blpage{90}
(\byear{2018})
{\href{https://arxiv.org/abs/1609.08373}{{arXiv:1609.08373}}}
{[gr-qc]}.
\doiurl{10.1016/j.dark.2017.12.005}
\end{barticle}
\endbibitem

\bibitem{Wright:2016ayu}
\begin{barticle}
\bauthor{\bsnm{Wright}, \binits{M.}}:
\batitle{{Conformal transformations in modified teleparallel theories of
  gravity revisited}}.
\bjtitle{Phys. Rev. D}
\bvolume{93}(\bissue{10}),
\bfpage{103002}
(\byear{2016})
{\href{https://arxiv.org/abs/1602.05764}{{arXiv:1602.05764}}}
{[gr-qc]}.
\doiurl{10.1103/PhysRevD.93.103002}
\end{barticle}
\endbibitem

\bibitem{Farrugia:2020fcu}
\begin{barticle}
\bauthor{\bsnm{Farrugia}, \binits{G.}},
\bauthor{\bsnm{Levi~Said}, \binits{J.}},
\bauthor{\bsnm{Finch}, \binits{A.}}:
\batitle{{Gravitoelectromagnetism, Solar System Tests, and Weak-Field Solutions
  in $f (T,B)$ Gravity with Observational Constraints}}.
\bjtitle{Universe}
\bvolume{6}(\bissue{2}),
\bfpage{34}
(\byear{2020})
{\href{https://arxiv.org/abs/2002.08183}{{arXiv:2002.08183}}}
{[gr-qc]}.
\doiurl{10.3390/universe6020034}
\end{barticle}
\endbibitem

\bibitem{Capozziello:2019msc}
\begin{barticle}
\bauthor{\bsnm{Capozziello}, \binits{S.}},
\bauthor{\bsnm{Capriolo}, \binits{M.}},
\bauthor{\bsnm{Caso}, \binits{L.}}:
\batitle{{Weak field limit and gravitational waves in $f(T,B)$ teleparallel
  gravity}}.
\bjtitle{Eur. Phys. J. C}
\bvolume{80}(\bissue{2}),
\bfpage{156}
(\byear{2020})
{\href{https://arxiv.org/abs/1912.12469}{{arXiv:1912.12469}}}
{[gr-qc]}.
\doiurl{10.1140/epjc/s10052-020-7737-9}
\end{barticle}
\endbibitem

\bibitem{Escamilla-Rivera:2019ulu}
\begin{barticle}
\bauthor{\bsnm{Escamilla-Rivera}, \binits{C.}},
\bauthor{\bsnm{Levi~Said}, \binits{J.}}:
\batitle{{Cosmological viable models in $f(T,B)$ theory as solutions to the
  $H_0$ tension}}.
\bjtitle{Class. Quant. Grav.}
\bvolume{37}(\bissue{16}),
\bfpage{165002}
(\byear{2020})
{\href{https://arxiv.org/abs/1909.10328}{{arXiv:1909.10328}}}
{[gr-qc]}.
\doiurl{10.1088/1361-6382/ab939c}
\end{barticle}
\endbibitem

\bibitem{Kofinas:2014owa}
\begin{barticle}
\bauthor{\bsnm{Kofinas}, \binits{G.}},
\bauthor{\bsnm{Saridakis}, \binits{E.N.}}:
\batitle{{Teleparallel equivalent of Gauss-Bonnet gravity and its
  modifications}}.
\bjtitle{Phys. Rev. D}
\bvolume{90},
\bfpage{084044}
(\byear{2014})
{\href{https://arxiv.org/abs/1404.2249}{{arXiv:1404.2249}}}
{[gr-qc]}.
\doiurl{10.1103/PhysRevD.90.084044}
\end{barticle}
\endbibitem

\bibitem{Kofinas:2014daa}
\begin{barticle}
\bauthor{\bsnm{Kofinas}, \binits{G.}},
\bauthor{\bsnm{Saridakis}, \binits{E.N.}}:
\batitle{{Cosmological applications of $F(T,T_G)$ gravity}}.
\bjtitle{Phys. Rev. D}
\bvolume{90},
\bfpage{084045}
(\byear{2014})
{\href{https://arxiv.org/abs/1408.0107}{{arXiv:1408.0107}}}
{[gr-qc]}.
\doiurl{10.1103/PhysRevD.90.084045}
\end{barticle}
\endbibitem

\bibitem{delaCruz-Dombriz:2017lvj}
\begin{barticle}
\bauthor{\bparticle{de~la} \bsnm{Cruz-Dombriz}, \binits{A.}},
\bauthor{\bsnm{Farrugia}, \binits{G.}},
\bauthor{\bsnm{Said}, \binits{J.L.}},
\bauthor{\bsnm{Saez-Gomez}, \binits{D.}}:
\batitle{{Cosmological reconstructed solutions in extended teleparallel gravity
  theories with a teleparallel Gauss\textendash{}Bonnet term}}.
\bjtitle{Class. Quant. Grav.}
\bvolume{34}(\bissue{23}),
\bfpage{235011}
(\byear{2017})
{\href{https://arxiv.org/abs/1705.03867}{{arXiv:1705.03867}}}
{[gr-qc]}.
\doiurl{10.1088/1361-6382/aa93c8}
\end{barticle}
\endbibitem

\bibitem{delaCruz-Dombriz:2018nvt}
\begin{barticle}
\bauthor{\bparticle{de~la} \bsnm{Cruz-Dombriz}, \binits{A.}},
\bauthor{\bsnm{Farrugia}, \binits{G.}},
\bauthor{\bsnm{Said}, \binits{J.L.}},
\bauthor{\bsnm{S\'aez-Chill\'on~G\'omez}, \binits{D.}}:
\batitle{{Cosmological bouncing solutions in extended teleparallel gravity
  theories}}.
\bjtitle{Phys. Rev. D}
\bvolume{97}(\bissue{10}),
\bfpage{104040}
(\byear{2018})
{\href{https://arxiv.org/abs/1801.10085}{{arXiv:1801.10085}}}
{[gr-qc]}.
\doiurl{10.1103/PhysRevD.97.104040}
\end{barticle}
\endbibitem

\bibitem{Bahamonde:2019ipm}
\begin{barticle}
\bauthor{\bsnm{Bahamonde}, \binits{S.}},
\bauthor{\bsnm{Dialektopoulos}, \binits{K.F.}},
\bauthor{\bsnm{Gakis}, \binits{V.}},
\bauthor{\bsnm{Levi~Said}, \binits{J.}}:
\batitle{{Reviving Horndeski theory using teleparallel gravity after
  GW170817}}.
\bjtitle{Phys. Rev. D}
\bvolume{101}(\bissue{8}),
\bfpage{084060}
(\byear{2020})
{\href{https://arxiv.org/abs/1907.10057}{{arXiv:1907.10057}}}
{[gr-qc]}.
\doiurl{10.1103/PhysRevD.101.084060}
\end{barticle}
\endbibitem

\bibitem{Bahamonde:2020cfv}
\begin{barticle}
\bauthor{\bsnm{Bahamonde}, \binits{S.}},
\bauthor{\bsnm{Dialektopoulos}, \binits{K.F.}},
\bauthor{\bsnm{Hohmann}, \binits{M.}},
\bauthor{\bsnm{Levi~Said}, \binits{J.}}:
\batitle{{Post-Newtonian limit of Teleparallel Horndeski gravity}}.
\bjtitle{Class. Quant. Grav.}
\bvolume{38}(\bissue{2}),
\bfpage{025006}
(\byear{2020})
{\href{https://arxiv.org/abs/2003.11554}{{arXiv:2003.11554}}}
{[gr-qc]}.
\doiurl{10.1088/1361-6382/abc441}
\end{barticle}
\endbibitem

\bibitem{Hohmann:2019nat}
\begin{barticle}
\bauthor{\bsnm{Hohmann}, \binits{M.}},
\bauthor{\bsnm{J\"arv}, \binits{L.}},
\bauthor{\bsnm{Kr\v{s}\v{s}\'ak}, \binits{M.}},
\bauthor{\bsnm{Pfeifer}, \binits{C.}}:
\batitle{{Modified teleparallel theories of gravity in symmetric spacetimes}}.
\bjtitle{Phys. Rev. D}
\bvolume{100}(\bissue{8}),
\bfpage{084002}
(\byear{2019})
{\href{https://arxiv.org/abs/1901.05472}{{arXiv:1901.05472}}}
{[gr-qc]}.
\doiurl{10.1103/PhysRevD.100.084002}
\end{barticle}
\endbibitem

\bibitem{Hayashi:1979qx}
\begin{barticle}
\bauthor{\bsnm{Hayashi}, \binits{K.}},
\bauthor{\bsnm{Shirafuji}, \binits{T.}}:
\batitle{{New General Relativity}}.
\bjtitle{Phys. Rev. D}
\bvolume{19},
\bfpage{3524}--\blpage{3553}
(\byear{1979}).
\doiurl{10.1103/PhysRevD.19.3524}.
\bcomment{[Addendum: Phys.Rev.D 24, 3312--3314 (1982)]}
\end{barticle}
\endbibitem

\bibitem{Chandrasekhar:1984siy}
\begin{barticle}
\bauthor{\bsnm{Chandrasekhar}, \binits{S.}}:
\batitle{{The Mathematical Theory of Black Holes}}.
\bjtitle{Fundam. Theor. Phys.}
\bvolume{9},
\bfpage{5}--\blpage{26}
(\byear{1984}).
\doiurl{10.1007/978-94-009-6469-3_2}
\end{barticle}
\endbibitem

\bibitem{DeFelice:2010aj}
\begin{barticle}
\bauthor{\bsnm{De~Felice}, \binits{A.}},
\bauthor{\bsnm{Tsujikawa}, \binits{S.}}:
\batitle{{f(R) theories}}.
\bjtitle{Living Rev. Rel.}
\bvolume{13},
\bfpage{3}
(\byear{2010})
{\href{https://arxiv.org/abs/1002.4928}{{arXiv:1002.4928}}}
{[gr-qc]}.
\doiurl{10.12942/lrr-2010-3}
\end{barticle}
\endbibitem

\bibitem{RezaeiAkbarieh:2018ijw}
\begin{barticle}
\bauthor{\bsnm{Rezaei~Akbarieh}, \binits{A.}},
\bauthor{\bsnm{Izadi}, \binits{Y.}}:
\batitle{{Tachyon Inflation in Teleparallel Gravity}}.
\bjtitle{Eur. Phys. J. C}
\bvolume{79}(\bissue{4}),
\bfpage{366}
(\byear{2019})
{\href{https://arxiv.org/abs/1812.06649}{{arXiv:1812.06649}}}
{[gr-qc]}.
\doiurl{10.1140/epjc/s10052-019-6819-z}
\end{barticle}
\endbibitem

\bibitem{Krssak:2015oua}
\begin{barticle}
\bauthor{\bsnm{Kr\v{s}\v{s}\'ak}, \binits{M.}},
\bauthor{\bsnm{Saridakis}, \binits{E.N.}}:
\batitle{{The covariant formulation of f(T) gravity}}.
\bjtitle{Class. Quant. Grav.}
\bvolume{33}(\bissue{11}),
\bfpage{115009}
(\byear{2016})
{\href{https://arxiv.org/abs/1510.08432}{{arXiv:1510.08432}}}
{[gr-qc]}.
\doiurl{10.1088/0264-9381/33/11/115009}
\end{barticle}
\endbibitem

\bibitem{Tamanini:2012hg}
\begin{barticle}
\bauthor{\bsnm{Tamanini}, \binits{N.}},
\bauthor{\bsnm{Boehmer}, \binits{C.G.}}:
\batitle{{Good and bad tetrads in f(T) gravity}}.
\bjtitle{Phys. Rev. D}
\bvolume{86},
\bfpage{044009}
(\byear{2012})
{\href{https://arxiv.org/abs/1204.4593}{{arXiv:1204.4593}}}
{[gr-qc]}.
\doiurl{10.1103/PhysRevD.86.044009}
\end{barticle}
\endbibitem

\bibitem{Jimenez:2001gg}
\begin{barticle}
\bauthor{\bsnm{Jimenez}, \binits{R.}},
\bauthor{\bsnm{Loeb}, \binits{A.}}:
\batitle{{Constraining cosmological parameters based on relative galaxy ages}}.
\bjtitle{Astrophys. J.}
\bvolume{573},
\bfpage{37}--\blpage{42}
(\byear{2002})
{\href{https://arxiv.org/abs/astro-ph/0106145}{{arXiv:astro-ph/0106145}}}
{[astro-ph]}.
\doiurl{10.1086/340549}
\end{barticle}
\endbibitem

\bibitem{2014RAA....14.1221Z}
\begin{barticle}
\bauthor{\bsnm{{Zhang}}, \binits{C.}},
\bauthor{\bsnm{{Zhang}}, \binits{H.}},
\bauthor{\bsnm{{Yuan}}, \binits{S.}},
\bauthor{\bsnm{{Liu}}, \binits{S.}},
\bauthor{\bsnm{{Zhang}}, \binits{T.-J.}},
\bauthor{\bsnm{{Sun}}, \binits{Y.-C.}}:
\batitle{{Four new observational H(z) data from luminous red galaxies in the
  Sloan Digital Sky Survey data release seven}}.
\bjtitle{Research in Astronomy and Astrophysics}
\bvolume{14}(\bissue{10}),
\bfpage{1221}--\blpage{1233}
(\byear{2014})
{\href{https://arxiv.org/abs/1207.4541}{{arXiv:1207.4541}}}
{[astro-ph.CO]}.
\doiurl{10.1088/1674-4527/14/10/002}
\end{barticle}
\endbibitem

\bibitem{Jimenez:2003iv}
\begin{barticle}
\bauthor{\bsnm{Jimenez}, \binits{R.}},
\bauthor{\bsnm{Verde}, \binits{L.}},
\bauthor{\bsnm{Treu}, \binits{T.}},
\bauthor{\bsnm{Stern}, \binits{D.}}:
\batitle{{Constraints on the equation of state of dark energy and the Hubble
  constant from stellar ages and the CMB}}.
\bjtitle{Astrophys. J.}
\bvolume{593},
\bfpage{622}--\blpage{629}
(\byear{2003})
{\href{https://arxiv.org/abs/astro-ph/0302560}{{arXiv:astro-ph/0302560}}}.
\doiurl{10.1086/376595}
\end{barticle}
\endbibitem

\bibitem{Moresco:2016mzx}
\begin{barticle}
\bauthor{\bsnm{Moresco}, \binits{M.}},
\bauthor{\bsnm{Pozzetti}, \binits{L.}},
\bauthor{\bsnm{Cimatti}, \binits{A.}},
\bauthor{\bsnm{Jimenez}, \binits{R.}},
\bauthor{\bsnm{Maraston}, \binits{C.}},
\bauthor{\bsnm{Verde}, \binits{L.}},
\bauthor{\bsnm{Thomas}, \binits{D.}},
\bauthor{\bsnm{Citro}, \binits{A.}},
\bauthor{\bsnm{Tojeiro}, \binits{R.}},
\bauthor{\bsnm{Wilkinson}, \binits{D.}}:
\batitle{{A 6\% measurement of the Hubble parameter at $z\sim0.45$: direct
  evidence of the epoch of cosmic re-acceleration}}.
\bjtitle{JCAP}
\bvolume{05},
\bfpage{014}
(\byear{2016})
{\href{https://arxiv.org/abs/1601.01701}{{arXiv:1601.01701}}}
{[astro-ph.CO]}.
\doiurl{10.1088/1475-7516/2016/05/014}
\end{barticle}
\endbibitem

\bibitem{Simon:2004tf}
\begin{barticle}
\bauthor{\bsnm{Simon}, \binits{J.}},
\bauthor{\bsnm{Verde}, \binits{L.}},
\bauthor{\bsnm{Jimenez}, \binits{R.}}:
\batitle{{Constraints on the redshift dependence of the dark energy
  potential}}.
\bjtitle{Phys. Rev. D}
\bvolume{71},
\bfpage{123001}
(\byear{2005})
{\href{https://arxiv.org/abs/astro-ph/0412269}{{arXiv:astro-ph/0412269}}}.
\doiurl{10.1103/PhysRevD.71.123001}
\end{barticle}
\endbibitem

\bibitem{2012JCAP...08..006M}
\begin{barticle}
\bauthor{\bsnm{{Moresco}}, \binits{M.}},
\bauthor{\bsnm{{Cimatti}}, \binits{A.}},
\bauthor{\bsnm{{Jimenez}}, \binits{R.}},
\bauthor{\bsnm{{Pozzetti}}, \binits{L.}},
\bauthor{\bsnm{{Zamorani}}, \binits{G.}},
\bauthor{\bsnm{{Bolzonella}}, \binits{M.}},
\bauthor{\bsnm{{Dunlop}}, \binits{J.}},
\bauthor{\bsnm{{Lamareille}}, \binits{F.}},
\bauthor{\bsnm{{Mignoli}}, \binits{M.}},
\bauthor{\bsnm{{Pearce}}, \binits{H.}},
\bauthor{\bsnm{{Rosati}}, \binits{P.}},
\bauthor{\bsnm{{Stern}}, \binits{D.}},
\bauthor{\bsnm{{Verde}}, \binits{L.}},
\bauthor{\bsnm{{Zucca}}, \binits{E.}},
\bauthor{\bsnm{{Carollo}}, \binits{C.M.}},
\bauthor{\bsnm{{Contini}}, \binits{T.}},
\bauthor{\bsnm{{Kneib}}, \binits{J.-P.}},
\bauthor{\bsnm{{Le F{\`e}vre}}, \binits{O.}},
\bauthor{\bsnm{{Lilly}}, \binits{S.J.}},
\bauthor{\bsnm{{Mainieri}}, \binits{V.}},
\bauthor{\bsnm{{Renzini}}, \binits{A.}},
\bauthor{\bsnm{{Scodeggio}}, \binits{M.}},
\bauthor{\bsnm{{Balestra}}, \binits{I.}},
\bauthor{\bsnm{{Gobat}}, \binits{R.}},
\bauthor{\bsnm{{McLure}}, \binits{R.}},
\bauthor{\bsnm{{Bardelli}}, \binits{S.}},
\bauthor{\bsnm{{Bongiorno}}, \binits{A.}},
\bauthor{\bsnm{{Caputi}}, \binits{K.}},
\bauthor{\bsnm{{Cucciati}}, \binits{O.}},
\bauthor{\bsnm{{de la Torre}}, \binits{S.}},
\bauthor{\bsnm{{de Ravel}}, \binits{L.}},
\bauthor{\bsnm{{Franzetti}}, \binits{P.}},
\bauthor{\bsnm{{Garilli}}, \binits{B.}},
\bauthor{\bsnm{{Iovino}}, \binits{A.}},
\bauthor{\bsnm{{Kampczyk}}, \binits{P.}},
\bauthor{\bsnm{{Knobel}}, \binits{C.}},
\bauthor{\bsnm{{Kova{\v{c}}}}, \binits{K.}},
\bauthor{\bsnm{{Le Borgne}}, \binits{J.-F.}},
\bauthor{\bsnm{{Le Brun}}, \binits{V.}},
\bauthor{\bsnm{{Maier}}, \binits{C.}},
\bauthor{\bsnm{{Pell{\'o}}}, \binits{R.}},
\bauthor{\bsnm{{Peng}}, \binits{Y.}},
\bauthor{\bsnm{{Perez-Montero}}, \binits{E.}},
\bauthor{\bsnm{{Presotto}}, \binits{V.}},
\bauthor{\bsnm{{Silverman}}, \binits{J.D.}},
\bauthor{\bsnm{{Tanaka}}, \binits{M.}},
\bauthor{\bsnm{{Tasca}}, \binits{L.A.M.}},
\bauthor{\bsnm{{Tresse}}, \binits{L.}},
\bauthor{\bsnm{{Vergani}}, \binits{D.}},
\bauthor{\bsnm{{Almaini}}, \binits{O.}},
\bauthor{\bsnm{{Barnes}}, \binits{L.}},
\bauthor{\bsnm{{Bordoloi}}, \binits{R.}},
\bauthor{\bsnm{{Bradshaw}}, \binits{E.}},
\bauthor{\bsnm{{Cappi}}, \binits{A.}},
\bauthor{\bsnm{{Chuter}}, \binits{R.}},
\bauthor{\bsnm{{Cirasuolo}}, \binits{M.}},
\bauthor{\bsnm{{Coppa}}, \binits{G.}},
\bauthor{\bsnm{{Diener}}, \binits{C.}},
\bauthor{\bsnm{{Foucaud}}, \binits{S.}},
\bauthor{\bsnm{{Hartley}}, \binits{W.}},
\bauthor{\bsnm{{Kamionkowski}}, \binits{M.}},
\bauthor{\bsnm{{Koekemoer}}, \binits{A.M.}},
\bauthor{\bsnm{{L{\'o}pez-Sanjuan}}, \binits{C.}},
\bauthor{\bsnm{{McCracken}}, \binits{H.J.}},
\bauthor{\bsnm{{Nair}}, \binits{P.}},
\bauthor{\bsnm{{Oesch}}, \binits{P.}},
\bauthor{\bsnm{{Stanford}}, \binits{A.}},
\bauthor{\bsnm{{Welikala}}, \binits{N.}}:
\batitle{{Improved constraints on the expansion rate of the Universe up to z
  \raisebox{-0.5ex}\textasciitilde 1.1 from the spectroscopic evolution of
  cosmic chronometers}}.
\bjtitle{JCAP}
\bvolume{2012}(\bissue{8}),
\bfpage{006}
(\byear{2012})
{\href{https://arxiv.org/abs/1201.3609}{{arXiv:1201.3609}}}
{[astro-ph.CO]}.
\doiurl{10.1088/1475-7516/2012/08/006}
\end{barticle}
\endbibitem

\bibitem{2010JCAP...02..008S}
\begin{barticle}
\bauthor{\bsnm{{Stern}}, \binits{D.}},
\bauthor{\bsnm{{Jimenez}}, \binits{R.}},
\bauthor{\bsnm{{Verde}}, \binits{L.}},
\bauthor{\bsnm{{Kamionkowski}}, \binits{M.}},
\bauthor{\bsnm{{Stanford}}, \binits{S.A.}}:
\batitle{{Cosmic chronometers: constraining the equation of state of dark
  energy. I: H(z) measurements}}.
\bjtitle{JCAP}
\bvolume{2010}(\bissue{2}),
\bfpage{008}
(\byear{2010})
{\href{https://arxiv.org/abs/0907.3149}{{arXiv:0907.3149}}}
{[astro-ph.CO]}.
\doiurl{10.1088/1475-7516/2010/02/008}
\end{barticle}
\endbibitem

\bibitem{Moresco:2015cya}
\begin{barticle}
\bauthor{\bsnm{Moresco}, \binits{M.}}:
\batitle{{Raising the bar: new constraints on the Hubble parameter with cosmic
  chronometers at z \ensuremath{\sim} 2}}.
\bjtitle{Mon. Not. Roy. Astron. Soc.}
\bvolume{450}(\bissue{1}),
\bfpage{16}--\blpage{20}
(\byear{2015})
{\href{https://arxiv.org/abs/1503.01116}{{arXiv:1503.01116}}}
{[astro-ph.CO]}.
\doiurl{10.1093/mnrasl/slv037}
\end{barticle}
\endbibitem

\bibitem{Gomez-Valent:2018hwc}
\begin{barticle}
\bauthor{\bsnm{Gómez-Valent}, \binits{A.}},
\bauthor{\bsnm{Amendola}, \binits{L.}}:
\batitle{{$H_0$ from cosmic chronometers and Type Ia supernovae, with Gaussian
  Processes and the novel Weighted Polynomial Regression method}}.
\bjtitle{JCAP}
\bvolume{04},
\bfpage{051}
(\byear{2018})
{\href{https://arxiv.org/abs/1802.01505}{{arXiv:1802.01505}}}
{[astro-ph.CO]}.
\doiurl{10.1088/1475-7516/2018/04/051}
\end{barticle}
\endbibitem

\bibitem{Lopez-Corredoira:2017zfl}
\begin{barticle}
\bauthor{\bsnm{Lopez-Corredoira}, \binits{M.}},
\bauthor{\bsnm{Vazdekis}, \binits{A.}},
\bauthor{\bsnm{Gutierrez}, \binits{C.M.}},
\bauthor{\bsnm{Castro-Rodriguez}, \binits{N.}}:
\batitle{{Stellar content of extremely red quiescent galaxies at z
  \ensuremath{>} 2}}.
\bjtitle{Astron. Astrophys.}
\bvolume{600},
\bfpage{91}
(\byear{2017})
{\href{https://arxiv.org/abs/1702.00380}{{arXiv:1702.00380}}}
{[astro-ph.GA]}.
\doiurl{10.1051/0004-6361/201629857}
\end{barticle}
\endbibitem

\bibitem{Lopez-Corredoira:2018tmn}
\begin{barticle}
\bauthor{\bsnm{Lopez-Corredoira}, \binits{M.}},
\bauthor{\bsnm{Vazdekis}, \binits{A.}}:
\batitle{{Impact of young stellar components on quiescent galaxies:
  deconstructing cosmic chronometers}}.
\bjtitle{Astron. Astrophys.}
\bvolume{614},
\bfpage{127}
(\byear{2018})
{\href{https://arxiv.org/abs/1802.09473}{{arXiv:1802.09473}}}
{[astro-ph.CO]}.
\doiurl{10.1051/0004-6361/201731647}
\end{barticle}
\endbibitem

\bibitem{Verde:2014qea}
\begin{barticle}
\bauthor{\bsnm{Verde}, \binits{L.}},
\bauthor{\bsnm{Protopapas}, \binits{P.}},
\bauthor{\bsnm{Jimenez}, \binits{R.}}:
\batitle{{The expansion rate of the intermediate Universe in light of Planck}}.
\bjtitle{Phys. Dark Univ.}
\bvolume{5-6},
\bfpage{307}--\blpage{314}
(\byear{2014})
{\href{https://arxiv.org/abs/1403.2181}{{arXiv:1403.2181}}}
{[astro-ph.CO]}.
\doiurl{10.1016/j.dark.2014.09.003}
\end{barticle}
\endbibitem

\bibitem{Scolnic:2017caz}
\begin{barticle}
\bauthor{\bsnm{Scolnic}, \binits{D.M.}}, \betal:
\batitle{{The complete light-curve sample of spectroscopically confirmed SNe Ia
  from Pan-STARRS1 and cosmological constraints from the combined Pantheon
  Sample}}.
\bjtitle{Astrophys. J.}
\bvolume{859}(\bissue{2}),
\bfpage{101}
(\byear{2018})
{\href{https://arxiv.org/abs/1710.00845}{{arXiv:1710.00845}}}
{[astro-ph.CO]}.
\doiurl{10.3847/1538-4357/aab9bb}
\end{barticle}
\endbibitem

\bibitem{2011ApJS..192....1C}
\begin{barticle}
\bauthor{\bsnm{{Conley}}, \binits{A.}},
\bauthor{\bsnm{{Guy}}, \binits{J.}},
\bauthor{\bsnm{{Sullivan}}, \binits{M.}},
\bauthor{\bsnm{{Regnault}}, \binits{N.}},
\bauthor{\bsnm{{Astier}}, \binits{P.}},
\bauthor{\bsnm{{Balland}}, \binits{C.}},
\bauthor{\bsnm{{Basa}}, \binits{S.}},
\bauthor{\bsnm{{Carlberg}}, \binits{R.G.}},
\bauthor{\bsnm{{Fouchez}}, \binits{D.}},
\bauthor{\bsnm{{Hardin}}, \binits{D.}},
\bauthor{\bsnm{{Hook}}, \binits{I.M.}},
\bauthor{\bsnm{{Howell}}, \binits{D.A.}},
\bauthor{\bsnm{{Pain}}, \binits{R.}},
\bauthor{\bsnm{{Palanque-Delabrouille}}, \binits{N.}},
\bauthor{\bsnm{{Perrett}}, \binits{K.M.}},
\bauthor{\bsnm{{Pritchet}}, \binits{C.J.}},
\bauthor{\bsnm{{Rich}}, \binits{J.}},
\bauthor{\bsnm{{Ruhlmann-Kleider}}, \binits{V.}},
\bauthor{\bsnm{{Balam}}, \binits{D.}},
\bauthor{\bsnm{{Baumont}}, \binits{S.}},
\bauthor{\bsnm{{Ellis}}, \binits{R.S.}},
\bauthor{\bsnm{{Fabbro}}, \binits{S.}},
\bauthor{\bsnm{{Fakhouri}}, \binits{H.K.}},
\bauthor{\bsnm{{Fourmanoit}}, \binits{N.}},
\bauthor{\bsnm{{Gonz{\'a}lez-Gait{\'a}n}}, \binits{S.}},
\bauthor{\bsnm{{Graham}}, \binits{M.L.}},
\bauthor{\bsnm{{Hudson}}, \binits{M.J.}},
\bauthor{\bsnm{{Hsiao}}, \binits{E.}},
\bauthor{\bsnm{{Kronborg}}, \binits{T.}},
\bauthor{\bsnm{{Lidman}}, \binits{C.}},
\bauthor{\bsnm{{Mourao}}, \binits{A.M.}},
\bauthor{\bsnm{{Neill}}, \binits{J.D.}},
\bauthor{\bsnm{{Perlmutter}}, \binits{S.}},
\bauthor{\bsnm{{Ripoche}}, \binits{P.}},
\bauthor{\bsnm{{Suzuki}}, \binits{N.}},
\bauthor{\bsnm{{Walker}}, \binits{E.S.}}:
\batitle{{Supernova Constraints and Systematic Uncertainties from the First
  Three Years of the Supernova Legacy Survey}}.
\bjtitle{The Astrophysical Journal}
\bvolume{192}(\bissue{1}),
\bfpage{1}
(\byear{2011})
{\href{https://arxiv.org/abs/1104.1443}{{arXiv:1104.1443}}}
{[astro-ph.CO]}.
\doiurl{10.1088/0067-0049/192/1/1}
\end{barticle}
\endbibitem

\bibitem{Ross:2014qpa}
\begin{barticle}
\bauthor{\bsnm{Ross}, \binits{A.J.}},
\bauthor{\bsnm{Samushia}, \binits{L.}},
\bauthor{\bsnm{Howlett}, \binits{C.}},
\bauthor{\bsnm{Percival}, \binits{W.J.}},
\bauthor{\bsnm{Burden}, \binits{A.}},
\bauthor{\bsnm{Manera}, \binits{M.}}:
\batitle{{The clustering of the SDSS DR7 main Galaxy sample \textendash{} I. A
  4 per cent distance measure at $z = 0.15$}}.
\bjtitle{Mon. Not. Roy. Astron. Soc.}
\bvolume{449}(\bissue{1}),
\bfpage{835}--\blpage{847}
(\byear{2015})
{\href{https://arxiv.org/abs/1409.3242}{{arXiv:1409.3242}}}
{[astro-ph.CO]}.
\doiurl{10.1093/mnras/stv154}
\end{barticle}
\endbibitem

\bibitem{2011MNRAS.416.3017B}
\begin{barticle}
\bauthor{\bsnm{{Beutler}}, \binits{F.}},
\bauthor{\bsnm{{Blake}}, \binits{C.}},
\bauthor{\bsnm{{Colless}}, \binits{M.}},
\bauthor{\bsnm{{Jones}}, \binits{D.H.}},
\bauthor{\bsnm{{Staveley-Smith}}, \binits{L.}},
\bauthor{\bsnm{{Campbell}}, \binits{L.}},
\bauthor{\bsnm{{Parker}}, \binits{Q.}},
\bauthor{\bsnm{{Saunders}}, \binits{W.}},
\bauthor{\bsnm{{Watson}}, \binits{F.}}:
\batitle{{The 6dF Galaxy Survey: baryon acoustic oscillations and the local
  Hubble constant}}.
\bjtitle{Monthly Notices of the Royal Astronomical Society}
\bvolume{416}(\bissue{4}),
\bfpage{3017}--\blpage{3032}
(\byear{2011})
{\href{https://arxiv.org/abs/1106.3366}{{arXiv:1106.3366}}}
{[astro-ph.CO]}.
\doiurl{10.1111/j.1365-2966.2011.19250.x}
\end{barticle}
\endbibitem

\bibitem{Bourboux:2017cbm}
\begin{barticle}
\bauthor{\bparticle{du~Mas~des} \bsnm{Bourboux}, \binits{H.}}, \betal:
\batitle{{Baryon acoustic oscillations from the complete SDSS-III
  Ly$\alpha$-quasar cross-correlation function at $z=2.4$}}.
\bjtitle{Astron. Astrophys.}
\bvolume{608},
\bfpage{130}
(\byear{2017})
{\href{https://arxiv.org/abs/1708.02225}{{arXiv:1708.02225}}}
{[astro-ph.CO]}.
\doiurl{10.1051/0004-6361/201731731}
\end{barticle}
\endbibitem

\bibitem{Zhao:2018gvb}
\begin{barticle}
\bauthor{\bsnm{Zhao}, \binits{G.-B.}}, \betal:
\batitle{{The clustering of the SDSS-IV extended Baryon Oscillation
  Spectroscopic Survey DR14 quasar sample: a tomographic measurement of cosmic
  structure growth and expansion rate based on optimal redshift weights}}.
\bjtitle{Mon. Not. Roy. Astron. Soc.}
\bvolume{482}(\bissue{3}),
\bfpage{3497}--\blpage{3513}
(\byear{2019})
{\href{https://arxiv.org/abs/1801.03043}{{arXiv:1801.03043}}}
{[astro-ph.CO]}.
\doiurl{10.1093/mnras/sty2845}
\end{barticle}
\endbibitem

\bibitem{Alam:2016hwk}
\begin{barticle}
\bauthor{\bsnm{Alam}, \binits{S.}}, \betal:
\batitle{{The clustering of galaxies in the completed SDSS-III Baryon
  Oscillation Spectroscopic Survey: cosmological analysis of the DR12 galaxy
  sample}}.
\bjtitle{Mon. Not. Roy. Astron. Soc.}
\bvolume{470}(\bissue{3}),
\bfpage{2617}--\blpage{2652}
(\byear{2017})
{\href{https://arxiv.org/abs/1607.03155}{{arXiv:1607.03155}}}
{[astro-ph.CO]}.
\doiurl{10.1093/mnras/stx721}
\end{barticle}
\endbibitem

\bibitem{2009ApJ...707..916F}
\begin{barticle}
\bauthor{\bsnm{{Fixsen}}, \binits{D.J.}}:
\batitle{{The Temperature of the Cosmic Microwave Background}}.
\bjtitle{The Astrophysical Journal}
\bvolume{707}(\bissue{2}),
\bfpage{916}--\blpage{920}
(\byear{2009})
{\href{https://arxiv.org/abs/0911.1955}{{arXiv:0911.1955}}}
{[astro-ph.CO]}.
\doiurl{10.1088/0004-637X/707/2/916}
\end{barticle}
\endbibitem

\bibitem{Freedman:2019jwv}
\begin{barticle}
\bauthor{\bsnm{Freedman}, \binits{W.L.}}, \betal:
\batitle{{The Carnegie-Chicago Hubble Program. VIII. An independent
  determination of the Hubble constant based on the Tip of the Red Giant
  Branch}}.
\bjtitle{Astrophys. J.}
\bvolume{882}(\bissue{1}),
\bfpage{34}
(\byear{2019})
{\href{https://arxiv.org/abs/1907.05922}{{arXiv:1907.05922}}}
{[astro-ph.CO]}.
\doiurl{10.3847/1538-4357/ab2f73}
\end{barticle}
\endbibitem

\bibitem{Akaike:1974}
\begin{barticle}
\bauthor{\bsnm{Akaike}, \binits{H.}}:
\batitle{A new look at the statistical model identification}.
\bjtitle{IEEE Transactions on Automatic Control}
\bvolume{19}(\bissue{6}),
\bfpage{716}--\blpage{723}
(\byear{1974}).
\doiurl{10.1109/TAC.1974.1100705}
\end{barticle}
\endbibitem

\bibitem{10.1214/aos/1176344136}
\begin{barticle}
\bauthor{\bsnm{Schwarz}, \binits{G.}}:
\batitle{{Estimating the Dimension of a Model}}.
\bjtitle{The Annals of Statistics}
\bvolume{6}(\bissue{2}),
\bfpage{461}--\blpage{464}
(\byear{1978}).
\doiurl{10.1214/aos/1176344136}
\end{barticle}
\endbibitem

\bibitem{Dvali:2000hr}
\begin{barticle}
\bauthor{\bsnm{Dvali}, \binits{G.R.}},
\bauthor{\bsnm{Gabadadze}, \binits{G.}},
\bauthor{\bsnm{Porrati}, \binits{M.}}:
\batitle{{4-D gravity on a brane in 5-D Minkowski space}}.
\bjtitle{Phys. Lett. B}
\bvolume{485},
\bfpage{208}--\blpage{214}
(\byear{2000})
{\href{https://arxiv.org/abs/hep-th/0005016}{{arXiv:hep-th/0005016}}}.
\doiurl{10.1016/S0370-2693(00)00669-9}
\end{barticle}
\endbibitem

\bibitem{Barcenas-Enriquez:2018ili}
\begin{barticle}
\bauthor{\bsnm{B\'arcenas-Enr\'\i{}quez}, \binits{G.}},
\bauthor{\bsnm{Escamilla-Rivera}, \binits{C.}},
\bauthor{\bsnm{Garcia-Aspeitia}, \binits{M.A.}}:
\batitle{{Cosmological analysis of a Dvali-Gabadadze-Porrati stable model with
  H(z) observations}}.
\bjtitle{Rev. Mex. Fis.}
\bvolume{64}(\bissue{6}),
\bfpage{584}--\blpage{589}
(\byear{2018})
{\href{https://arxiv.org/abs/1803.03283}{{arXiv:1803.03283}}}
{[gr-qc]}.
\doiurl{10.31349/RevMexFis.64.584}
\end{barticle}
\endbibitem

\bibitem{2018ApJ...855...89X}
\begin{barticle}
\bauthor{\bsnm{{Xu}}, \binits{B.}},
\bauthor{\bsnm{{Yu}}, \binits{H.}},
\bauthor{\bsnm{{Wu}}, \binits{P.}}:
\batitle{{Testing Viable f(T) Models with Current Observations}}.
\bjtitle{The Astrophysical Journal}
\bvolume{855}(\bissue{2}),
\bfpage{89}
(\byear{2018}).
\doiurl{10.3847/1538-4357/aaad12}
\end{barticle}
\endbibitem

\bibitem{Wang:2020zfv}
\begin{barticle}
\bauthor{\bsnm{Wang}, \binits{D.}},
\bauthor{\bsnm{Mota}, \binits{D.}}:
\batitle{{Can $f(T)$ gravity resolve the $H_0$ tension?}}
\bjtitle{Phys. Rev. D}
\bvolume{102}(\bissue{6}),
\bfpage{063530}
(\byear{2020})
{\href{https://arxiv.org/abs/2003.10095}{{arXiv:2003.10095}}}
{[astro-ph.CO]}.
\doiurl{10.1103/PhysRevD.102.063530}
\end{barticle}
\endbibitem

\bibitem{Linder:2009jz}
\begin{barticle}
\bauthor{\bsnm{Linder}, \binits{E.V.}}:
\batitle{{Exponential Gravity}}.
\bjtitle{Phys. Rev. D}
\bvolume{80},
\bfpage{123528}
(\byear{2009})
{\href{https://arxiv.org/abs/0905.2962}{{arXiv:0905.2962}}}
{[astro-ph.CO]}.
\doiurl{10.1103/PhysRevD.80.123528}
\end{barticle}
\endbibitem

\bibitem{Bamba:2010wb}
\begin{barticle}
\bauthor{\bsnm{Bamba}, \binits{K.}},
\bauthor{\bsnm{Geng}, \binits{C.-Q.}},
\bauthor{\bsnm{Lee}, \binits{C.-C.}},
\bauthor{\bsnm{Luo}, \binits{L.-W.}}:
\batitle{{Equation of state for dark energy in $f(T)$ gravity}}.
\bjtitle{JCAP}
\bvolume{01},
\bfpage{021}
(\byear{2011})
{\href{https://arxiv.org/abs/1011.0508}{{arXiv:1011.0508}}}
{[astro-ph.CO]}.
\doiurl{10.1088/1475-7516/2011/01/021}
\end{barticle}
\endbibitem

\bibitem{Deffayet:2000uy}
\begin{barticle}
\bauthor{\bsnm{Deffayet}, \binits{C.}}:
\batitle{{Cosmology on a brane in Minkowski bulk}}.
\bjtitle{Phys. Lett. B}
\bvolume{502},
\bfpage{199}--\blpage{208}
(\byear{2001})
{\href{https://arxiv.org/abs/hep-th/0010186}{{arXiv:hep-th/0010186}}}.
\doiurl{10.1016/S0370-2693(01)00160-5}
\end{barticle}
\endbibitem

\bibitem{Wu:2010av}
\begin{barticle}
\bauthor{\bsnm{Wu}, \binits{P.}},
\bauthor{\bsnm{Yu}, \binits{H.W.}}:
\batitle{{$f(T)$ models with phantom divide line crossing}}.
\bjtitle{Eur. Phys. J. C}
\bvolume{71},
\bfpage{1552}
(\byear{2011})
{\href{https://arxiv.org/abs/1008.3669}{{arXiv:1008.3669}}}
{[gr-qc]}.
\doiurl{10.1140/epjc/s10052-011-1552-2}
\end{barticle}
\endbibitem

\end{thebibliography}


\end{document}